\documentclass[preprintnumbers,preprint,aps]{revtex4}
\usepackage{graphicx,epsfig}
\input axodraw.sty
\usepackage{amsfonts,amssymb,amsmath,amssymb}
 \usepackage{natbib}
\usepackage[dvips]{color}

\newcommand{\ba}{\begin{array}}
\newcommand{\ea}{\end{array}}
\newcommand{\beq}{\begin{equation}}
\newcommand{\eeq}{\end{equation}}

\def\pbpg{p_b\cdot p_g}
\def\ptpg{p_t\cdot p_g}
\def\ptpb{p_t\cdot p_b}

\def\pbn3{p_b\cdot n_3}
\def\pgn3{p_g\cdot n_3}

\def\be{\begin{equation}}
\def\ee{\end{equation}}
\def\bt{\begin{table}}
\def\et{\end{table}}
\def\bc{\begin{center}}
\def\ec{\end{center}}
\def\bi{\begin{itemize}}
\def\ei{\end{itemize}}
\def\bea{\begin{eqnarray}}
\def\eea{\end{eqnarray}}
\def\beas{\begin{eqnarray*}}
\def\eeas{\end{eqnarray*}}

\def\N0{\widetilde{\chi}^0}

\def\fra{\mathrm{f_{1R}}}
\def\fla{\mathrm{f_{1L}}}
\def\flb{\mathrm{f_{2L}}}
\def\frb{\mathrm{f_{2R}}}

\def\shat{\hat{s}}

\def\sbg{\mathcal S_{bg}}
\def\stg{\mathcal S_{tg}}
\def\stb{\mathcal S_{tb}}
\def\sbn{\mathcal S_{bn}}
\def\sgn{\mathcal S_{gn}}

\begin{document} 
\vspace*{1cm}
\title{\boldmath 
Probing anomalous $tbW$ couplings in single-top production using
top polarization at the Large Hadron Collider }
\bigskip
\author{Saurabh D. Rindani\footnote{saurabh@prl.res.in} and Pankaj Sharma\footnote{pankajs@prl.res.in}}
\smallskip
\affiliation{Theoretical Physics Division, 
Physical Research Laboratory, 
Navrangpura, Ahmedabad 380 009, India \vskip 1.0 truecm}

\begin{abstract}
\vskip 0.5 truecm 
We study the sensitivity of the Large Hadron Collider (LHC) to anomalous $tbW$ couplings in single-top production in association with a $W^-$ boson 
followed by semileptonic decay of the top. We calculate top polarization and the effects of these anomalous couplings to it at two centre-of-mass 
(cm) energies of 7 TeV and 14 TeV. As a measure of top polarization, we look at various laboratory frame distributions of its decay products, viz., 
lepton angular and energy distributions and $b$-quark angular distributions, without requiring reconstruction of the rest frame of the top,
and study the effect of anomalous couplings on these distributions. We construct certain asymmetries  
to study the sensitivity of these distributions to anomalous $tbW$ couplings. We find that 1$\sigma$ limits on real and imaginary parts of the
dominant anomalous coupling $\frb$ which may be obtained by utilizing these asymmetries at the LHC with cm energy of 14 TeV and an integrated
luminosity of 10 fb$^{-1}$ will be significantly better than the expectations from direct measurements of cross sections and some other variables at the
LHC and over an order of magnitude better than the indirect limits.
\end{abstract}

 \maketitle
\section{Introduction}
Since the discovery of the top quark at Tevatron \cite{Abe:1994xt}, its properties have been studied extensively, in particular its mass \cite{:2007bxa}. 
With a mass $m_t= 172.6\pm1.4$ GeV  close to the electroweak (EW) scale, the top quark is widely considered to be a key to the understanding of
 the mechanism of electroweak symmetry breaking. Owing to its large mass, its decay width $\Gamma_t=1.3$ GeV is very small, leading to an 
extremely short life time, $\tau_t=1/\Gamma_t\simeq 5\times 10^{-25}$ s.  This is an order of magnitude smaller than the hadronization time scale 
$\tau_{had}\simeq 1/\Lambda_{QCD}\sim 3\times 10^{-24}$ s, and hence it decays before any non-perturbative effects can spoil its spin 
information. As a consequence, its polarization can be determined through the angular and/or energy distributions of its decay products \cite{Bigi:1986jk}. 
The degree of polarization of the top quark may be a probe of any new physics responsible for its production. Top-quark polarization has been suggested for the study 
of new physics models in top-pair as well as single-top production at hadron colliders \cite{Huitu:2010ad,Arai:2010ci,Godbole:2010kr, Choudhury:2010cd, Berger:2011hn}.

At the Large Hadron Collider (LHC), top quarks will be produced mainly in pairs dominantly through gluon fusion whose contribution to the cross section 
is 90\%, while quark-antiquark annihilation contributes 10\%. Both these production mechanisms proceed mainly through QCD interactions. 
Since gluon couplings are chirality-conserving, the polarization of top quarks can arise only through  the $Z$-exchange contribution to $q\bar
q$ annihilation channel, and is expected to be very small. However,
single-top production can also occur \cite{Heinson:1996zm,Stelzer:1998ni,Belyaev:1998dn,
Boos:1999dd,Tait:1999cf,Espriu:2001vj,Espriu:2002wx,Tait:2000sh,Chen:2005vr,AguilarSaavedra:2011ct},
and has already been seen \cite{Chatrchyan:2011vp}. Since it proceeds via weak interactions, 
top quarks will have large 
polarization \cite{Tait:2000sh,Espriu:2002wx}. At LHC energies, single-top quark events in the SM are expected to be produced 
via  a) the $t$-channel ($bq\rightarrow tq^\prime$) process, b) the $s$-channel process ($q\bar q^\prime\rightarrow t\bar{b}$) 
and c) the $tW$ associated production process ($bg\rightarrow tW^-$) \cite{Tait:1999cf}. In all these processes, there is at least one chiral vertex which gives
rise to large top polarization. The three processes are completely different kinematically and can be separated from each other. 
In this work, we study in detail the effect of anomalous $tbW$ 
vertex on top polarization in associated production of the top quark with a $W$ boson.

Among all top couplings to gauge and Higgs bosons, the $tbW$ vertex deserves special attention since the top quark is expected to decay almost
completely via this interaction. However, in several extensions of SM, as for example supersymmetry and models of dynamical symmetry breaking, 
sizable deviations are possible from the SM predictions and also new decays of top quarks are possible. These deviations of the $tbW$ vertex 
may be observed in top decays. Single-top production is another source to study deviation in the $tbW$ vertex. 
For on-shell top, bottom and $W$, the most general effective vertices for the $tbW$ interaction up to dimension five can be 
written as \cite{AguilarSaavedra:2006fy}
\begin{equation}
 V_{t\to bW^+} =\frac{-g}{\sqrt{2}}V_{tb}\left[\gamma^{\mu}
(\fla P_{L}+\fra P_{R})-
\frac{i \sigma^{\mu \nu}}{m_W}(p_t -p_b)_{\nu}(\flb P_{L}+\frb
P_{R})\right] \label{anomaloustbW1}
\end{equation}
for the decay $t\to bW^+$, and
\begin{equation}
 V_{b\to tW^-} =\frac{-g}{\sqrt{2}}V_{tb}^* \left[\gamma^{\mu}
(\mathrm {f_{1L}^*} P_{L}+\mathrm {f_{1R}^*} P_{R})-
\frac{i \sigma^{\mu \nu}}{m_W}(p_t -p_b)_{\nu}(\mathrm {f_{2R}^*} P_{L}+\mathrm {f_{2L}^*}
P_{R})\right] \label{anomaloustbW2}
\end{equation}
for $tW^-$ production from a virtual $b$, where $V_{tb}$ is the Cabibbo-Kobayashi-Maskawa matrix element, and
$\mathrm{f_{1L},~f_{2L}}$, $\mathrm{f_{1R},~f_{2R}}$ are couplings. 
In the SM, $\fla=1$ and $\fra=\flb=\frb=0$. We have assumed all the 
anomalous couplings to be complex and treat real and imaginary parts of these couplings as independent parameters. We assume CP to be conserved 
in this analysis. Various extensions of the SM would have specific predictions for these anomalous couplings. For example, the contributions to 
these form factors in two Higgs doublet model (2HDM), minimal supersymmetric standard model (MSSM) and top-color assisted 
Technicolor model (TC2) have been evaluated in Ref. \cite{Bernreuther:2008us}. 

 Tevatron provides the only existing direct limits on anomalous $tbW$ couplings through $W$ polarization measurements.  
Recently, CDF II with 2.7 fb$^{-1}$ of collected data reported 
results for the longitudinal and right-handed helicity fractions 
of the $W$ boson in semileptonic top-decays \cite{WpolCDF}. 
D0 also reported results on $W$ helicity fractions using a 
combination of semileptonic and 
dilepton decay channels \cite{WpolD0}.
These results on $W$ polarization measurements led to a limit of 
[$-0.3$, 0.3] on $\frb$ at the 95\% confidence level (CL) 
\cite{Chen:2005vr} in single-top production. The CMS and ATLAS
collaborations have also reported the $W$ helicity fractions with early LHC data \cite{WpolLHC}. 
In Ref. \cite{AguilarSaavedra:2011ct}, the authors have used the recent top quark decay asymmetries from ATLAS and the $t$-channel single-top cross 
section from CMS to put the limits on $tbW$ couplings. They find that despite the small statistics available, the early LHC limits 
of [$-0.6$, 0.3] on $\frb$ are not too far from the Tevatron limits.

Apart from direct measurements at the LHC and Tevatron, there are stringent indirect constraints coming from 
low-energy measurements of anomalous $tbW$ couplings. The measured rate of $b\rightarrow s\gamma$ puts stringent 
constraints on the couplings $\fra$ and $\flb$ of about $4\times10^{-3}$, since their contributions to $B$-meson decay get 
an enhancement factor of $m_t/m_b$ \cite{Cho:1993zb,Larios:1999au,Burdman:1999fw,Grzadkowski:2008mf}. The bound on the
anomalous coupling $\frb$ is very weak, [$-0.15$, 0.57] at 95 \% CL. These bounds are obtained by taking one coupling to be 
non-zero at a time. However if one allows all couplings to be non-zero simultaneously, there is a possibility of cancellations 
among contributions of different couplings and the limits on these couplings could be very different.

Single-top production in association with a $W^-$ boson can be used to probe the size and nature of 
the $tbW$ couplings. Apart from the cross section, the angular distribution of the top, and even the
polarization of the top would give additional information enabling the determination of the $tbW$ coupling. 
Here we focus on the polarization of the top produced in the $tW$ associated production process. 
The most direct way to determine top polarization is by measuring the angular distribution of 
its decay products in its rest frame. However, at the LHC reconstructing the top rest frame 
is difficult and will result in large systematic uncertainties. In this paper, we show how the decay lepton angular distributions in the
laboratory (lab) frame can be a useful probe of top polarization and hence of anomalous $tbW^-$ couplings. 

In this context it is interesting to recall that the angular distribution of the charged lepton is independent of anomalous 
$tbW$ couplings in the decay of the top quark, to linear order in the
anomalous couplings (the ``decoupling theorem'') 
\cite{decoupling,Godbole:2006tq}. Hence, the lepton angular distribution may be considered to be a pure and robust probe of any new physics involved 
in the production of the top quark. In this analysis, we study the angular distribution of the charged lepton without making the 
approximation that anomalous couplings are small and include higher-order terms in anomalous couplings. We also study the distributions
to linear order in the anomalous couplings in the region of validity of the decoupling theorem.
In this way, we investigate the range of anomalous couplings for which the decoupling theorem is valid.
We find that the azimuthal distribution of the lepton is sensitive to top polarization and can be used 
to probe the form factor $\frb$. The use of laboratory-frame charged-lepton azimuthal distribution for the measurement of top polarization in $t\bar t$
production at the LHC in a certain class of $Z'$ models was demonstrated in \cite{Godbole:2010kr}. 

While charged-lepton angular distributions, at least for small values of anomalous couplings, would be a direct measure of top polarization, and
hence would probe anomalous $tbW$ couplings in $tW^-$ production, other kinematic distributions could also be used to study anomalous $tbW$
couplings. Thus, the charged-lepton energy distribution and the decay-$b$ angular distribution, which do not obey the decoupling theorem, would
depend on top polarization as well as anomalous couplings in top decay. Thus they would get contributions from $tbW$ couplings in production as
well as in decay. We have thus also studied the charged-lepton energy distribution and the $b$-quark angular distribution as probes of
anomalous $tbW$ couplings. In all cases, we obtain analytical expressions for the parton-level distributions at leading order in the
couplings.

The effects of top polarization in $tW$ and $tH^-$ production have been studied previously in \cite{Beccaria:2004xk,Huitu:2010ad}. 
Ref. \cite{Beccaria:2004xk} included the effects of one-loop EW SUSY corrections.
It did not however include top decay. Top polarization in different modes of single-top production 
has also been studied in \cite{Espriu:2001vj,Espriu:2002wx}, where
spin-sensitive variables were 
used to analyze effective left- and right-handed couplings of the top coming from physics beyond the SM.

The rest of the paper is organized as follows. In the next section we discuss the cross section for single-top production in association with
a $W^-$, the corresponding top angular distribution, 
as well as top polarization in the process, obtaining analytical expressions and evaluating them numerically. In Sec.
III, we obtain expressions for charged-lepton angular and energy distributions, and in Sec. IV we obtain the $b$-quark
angular distribution. In case of each distribution, we evaluate an asymmetry as a variable sensitive to $tbW$ anomalous couplings. Sec. V
contains an analysis of the sensitivities of the various asymmetries to the anomalous couplings. In Sec. VI we discuss briefly the backgrounds and
non-leading corrections to our signal process, and conclude in Sec. VII. The Appendix contains certain lengthy mathematical expressions.

\section{\boldmath Single-top production in association with a $W$ boson}

As stated earlier, single-top production at hadron colliders occurs through three different modes. All these three modes are 
distinct in terms of initial and final states and are in principle separately measurable. The $tW^-$ mode of 
single-top production is distinct from other two modes in the sense that it is affected by the new physics only in the
$tbW$ vertex while, in other modes, there may be exotic scalars or gauge bosons which can give additional 
contributions to the process. The $tW^-$ mode of single-top production has been studied in detail in Ref. \cite{Tait:1999cf}.
\begin{figure}[h]
\begin{center}
\begin{picture}(800,130)(0,0)
\Gluon(100,100)(140,60){5}{5}
\ArrowLine(100,20)(140,60)
\Vertex(140,60){1}
\ArrowLine(140,60)(190,60)
\Vertex(190,60){6}
\Photon(190,60)(230,100){5}{5}
\ArrowLine(190,60)(230,20)
\put(75,85){$g (p_g)$}
\put(75,35){$b (p_b)$}
\put(235,85){$W^- $}
\put(225,35){$t (p_t,\lambda_t)$}
\put(160, 45){$b$}
\put(150,00){$(a)$}
\Gluon(310,100)(350,80){5}{4}
\ArrowLine(350,80)(390,100)
\Vertex(350,80){1}
\ArrowLine(350,50)(350,80)
\Vertex(350,50){6}
\Photon(350,50)(390,20){5}{5}
\ArrowLine(310,20)(350,50)
\put(285,85){$g (p_g)$}
\put(285,35){$b (p_b)$}
\put(390,35){$W^- $}
\put(390,85){$t (p_t,\lambda_t)$}
\put(360,60){$t$}
\put(350,00){$(b)$}
\end{picture}
\caption{ Feynman diagrams contributing to associated $tW^-$ production at the LHC.}\label{feyngraph1}
\end{center}
\end{figure}
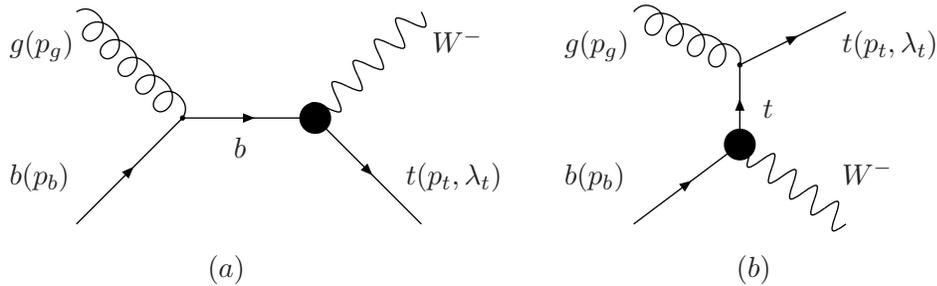
\begin{figure}[h]
\begin{center}
\begin{picture}(800,130)(0,0)
\ArrowLine(120,60)(190,60)
\Vertex(190,60){6}
\Photon(190,60)(250,60){5}{5}
\ArrowLine(190,60)(190,00)
\ArrowLine(250,60)(320,60)
\ArrowLine(250,120)(250,60)
\put(215,70){$W^+$}
\put(285,70){$\nu_{\ell^+}$}
\put(260,110){$\ell^+$}
\put(200,15){$b$}
\put(120, 45){$t$}
\end{picture}
\caption{ Anomalous $tbW$ couplings in top decay}\label{feyngraph2}
\end{center}
\end{figure}
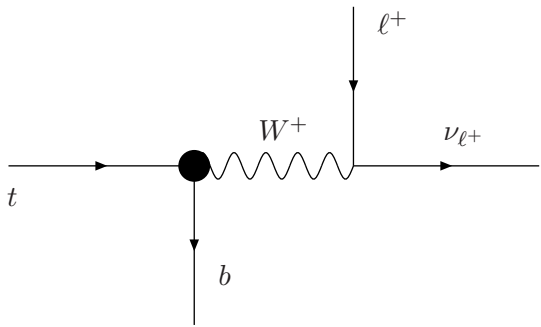
At the parton level, the $tW^-$ production proceeds through a gluon and a bottom quark each coming from a proton  and 
gets contribution from two diagrams. Feynman diagrams  for the process $g(p_g)b(p_b)\rightarrow t(p_t,\lambda_t)W^-$ 
are shown in Fig. \ref{feyngraph1}, where $\lambda_t=\pm1$ represent the
top helicity and the blobs denote effective
$tbW$ vertices, including anomalous couplings, in the production process. 
These couplings are also involved in the decay of the top as shown in Fig. \ref{feyngraph2}.

We obtain analytical expressions for the spin density matrix for $tW$ production including anomalous couplings. Use was made of the analytic manipulation
program FORM \cite{Vermaseren:2000nd}. We find that at linear order, only the real part of the coupling $\frb$ gives significant contribution to the 
production density matrix, whereas contributions from all other
couplings are proportional to the mass of $b$ quark (which we neglect
consistently) and hence vanish in the limit of zero bottom
mass. To second order in anomalous couplings, other anomalous couplings do contribute, but we focus only on $\frb$, since its contribution, arising at
linear order, is dominant. Expressions for the spin density matrix
elements $\rho(\pm,\pm)$ and $\rho(\pm,\mp)$ for $tW$ production, where $\pm$ are the signs of
the top-quark helicity, are given in the Appendix.

\subsection{Production cross section}

After integrating the density matrix given in the Appendix over
the phase space, the diagonal elements of this integrated density
matrix, which we denote by $\sigma(+,+)$ and  $\sigma(-,-)$, are 
respectively the cross sections for the production of positive and negative helicity tops and $\sigma_{\rm{tot}}=\sigma(+,+)+\sigma(-,-)$ 
is the total cross section. 

For numerical calculations, we use the leading-order parton distribution function (PDF) sets of CTEQ6L \cite{cteq6}
with a factorization scale of $m_t=172.6$ GeV. We also evaluate the strong
coupling at the same scale, $\alpha_s(m_t)=0.1085$. We make use of the 
following values of other parameters: $M_W=80.403$ GeV, the electromagnetic coupling $\alpha_{em}(m_Z)=1/128$ and $\sin^2\theta_W=0.23$.
We set $\fla = 1$ and $V_{tb}=1$ in our calculations. We take only one coupling to be non-zero at a time in the analysis,
except in Sec.\ref{sens}.

\begin{figure}[htb]
\begin{center}
\includegraphics[angle=270,width=3.2in]{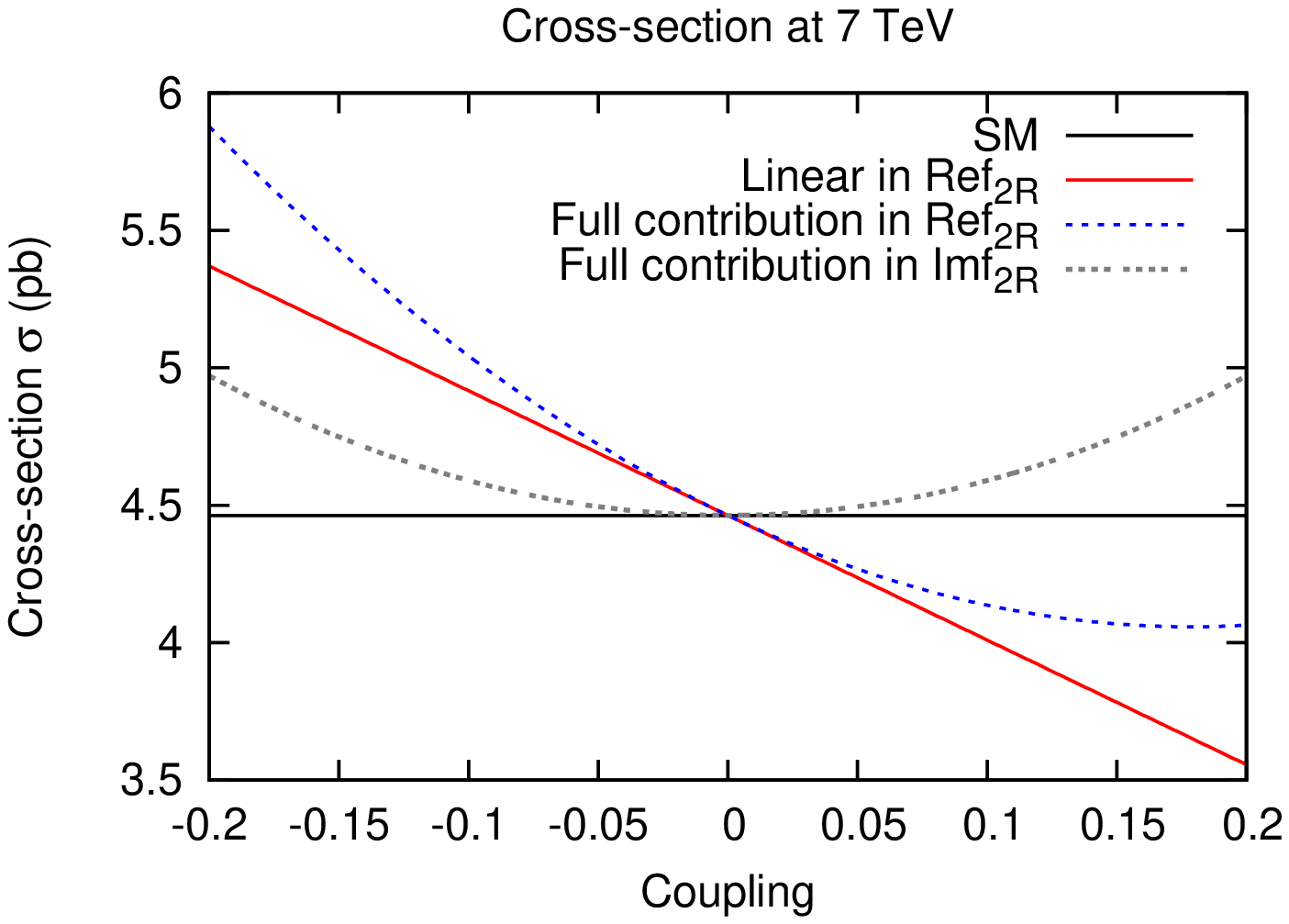} 
\includegraphics[angle=270,width=3.2in]{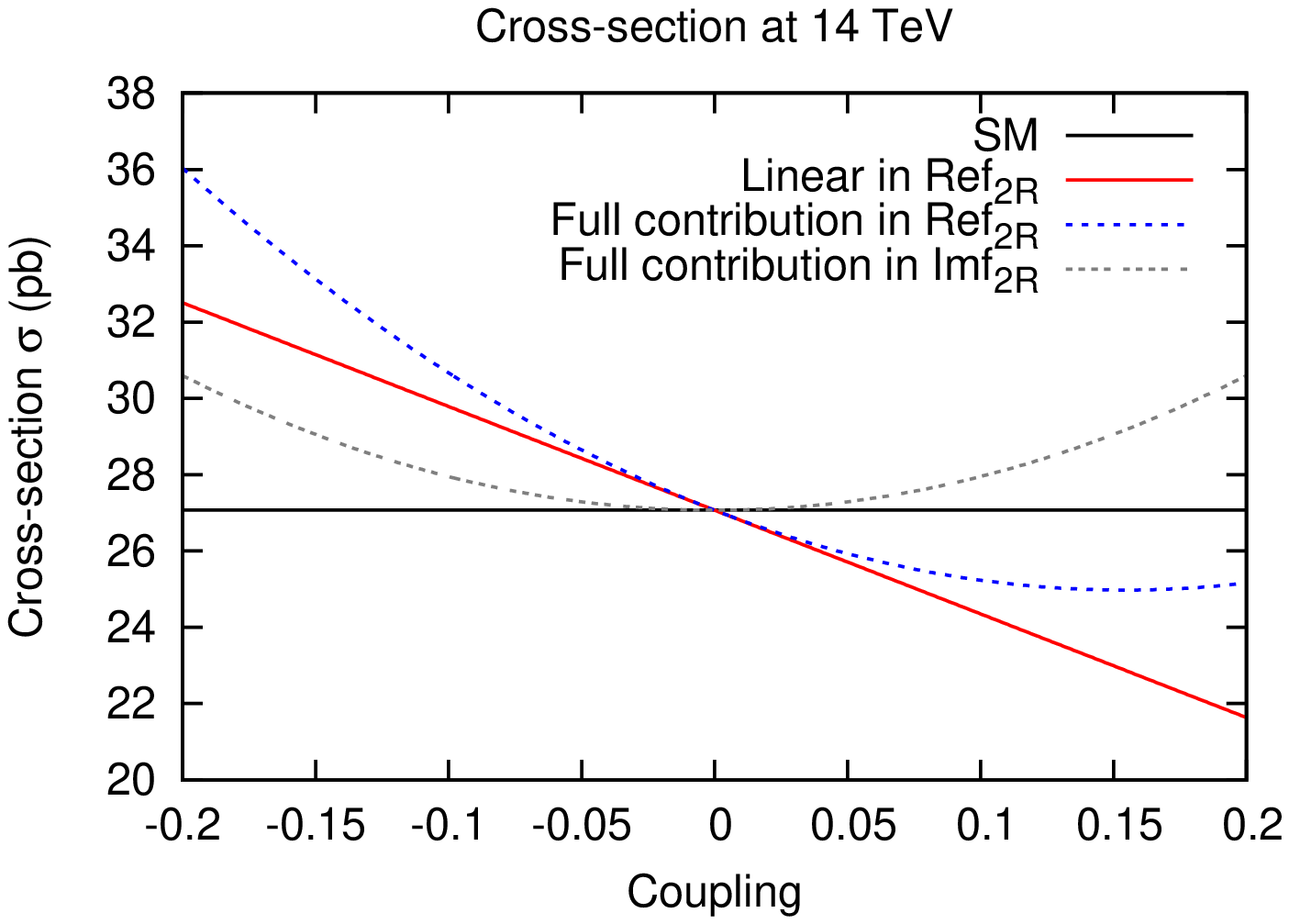} 
\caption{ The cross section for $tW^-$ production at the LHC for 
two different cm energies, 7 TeV (left) and 14 TeV (right), as a function of anomalous $tbW$ couplings.} 
\label{cs}
\end{center}
\end{figure}

In Fig. \ref{cs}, we show the cross section as a function of various anomalous $tbW$ couplings for two different 
values of centre-of-mass (cm) energies of 7 TeV and 14 TeV for which the LHC is planned to operate. 
We show contributions of anomalous couplings to the cross section at linear order as well as without the approximation.
From Fig. \ref{cs}, one can infer that the cross section is very sensitive to negative values of $\mathrm{Re}\frb$ 
The linear approximation is seen to be good for values of anomalous coupling ranging from $-0.05$ to 0.05. 

Since the cross section may receive large radiative corrections at the
LHC, we focus on using observables like asymmetries which are ratios of some
partial cross sections, and which are expected to be insensitive to such corrections.

\subsection{Top-angular distribution}

The angular distribution of the top quark would be modified by anomalous
couplings. Since the top quark is produced in a $2\to 2$ process, its azimuthal distribution is flat. 
We can study its polar distribution with the polar angle defined with respect 
to either of the beam directions as the $z$ axis. We find that the polar
distribution is sensitive to anomalous $tbW$ couplings.
\begin{figure}[htb]
\begin{center}
\includegraphics[angle=270,width=3.2in]{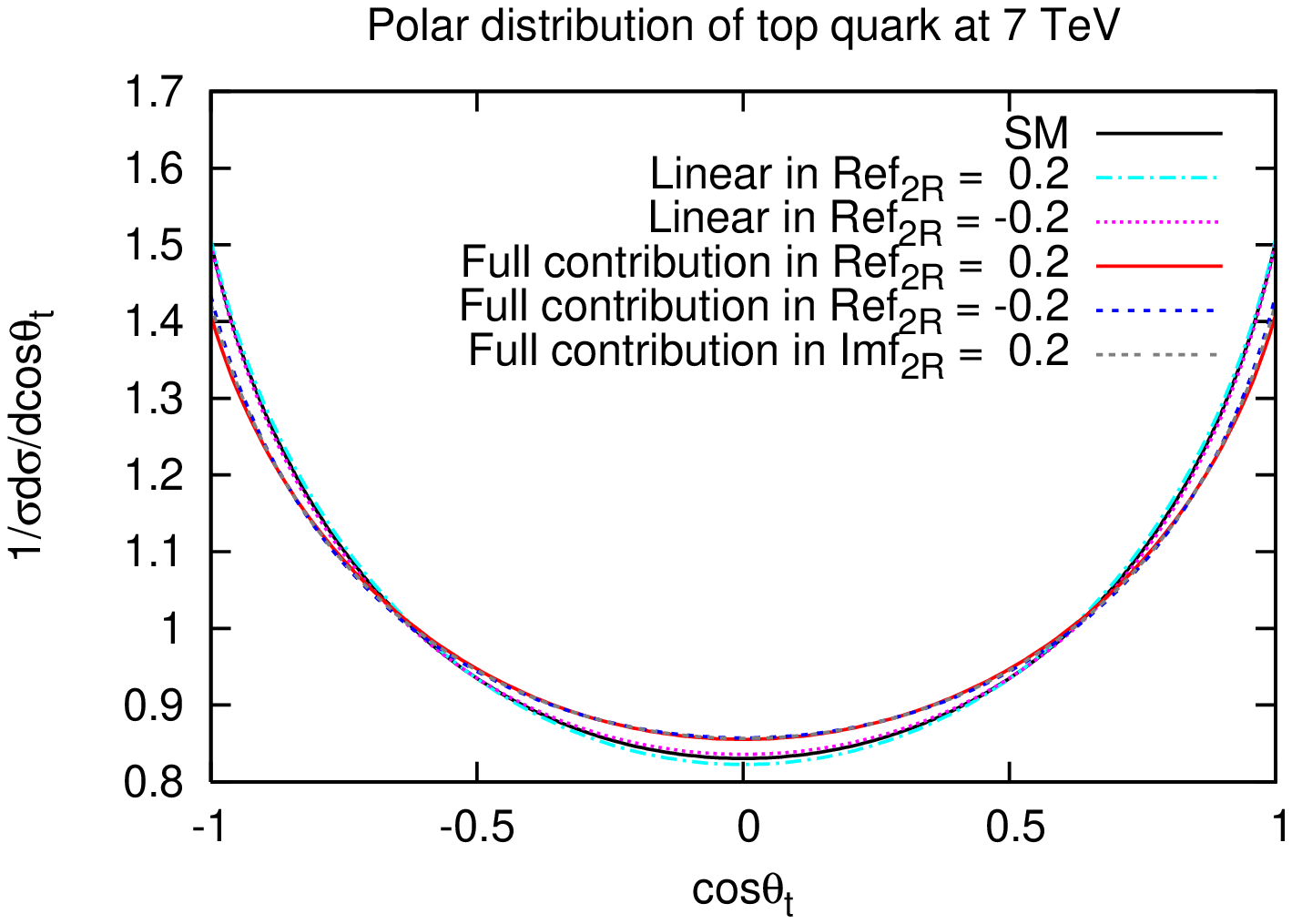} 
\includegraphics[angle=270,width=3.2in]{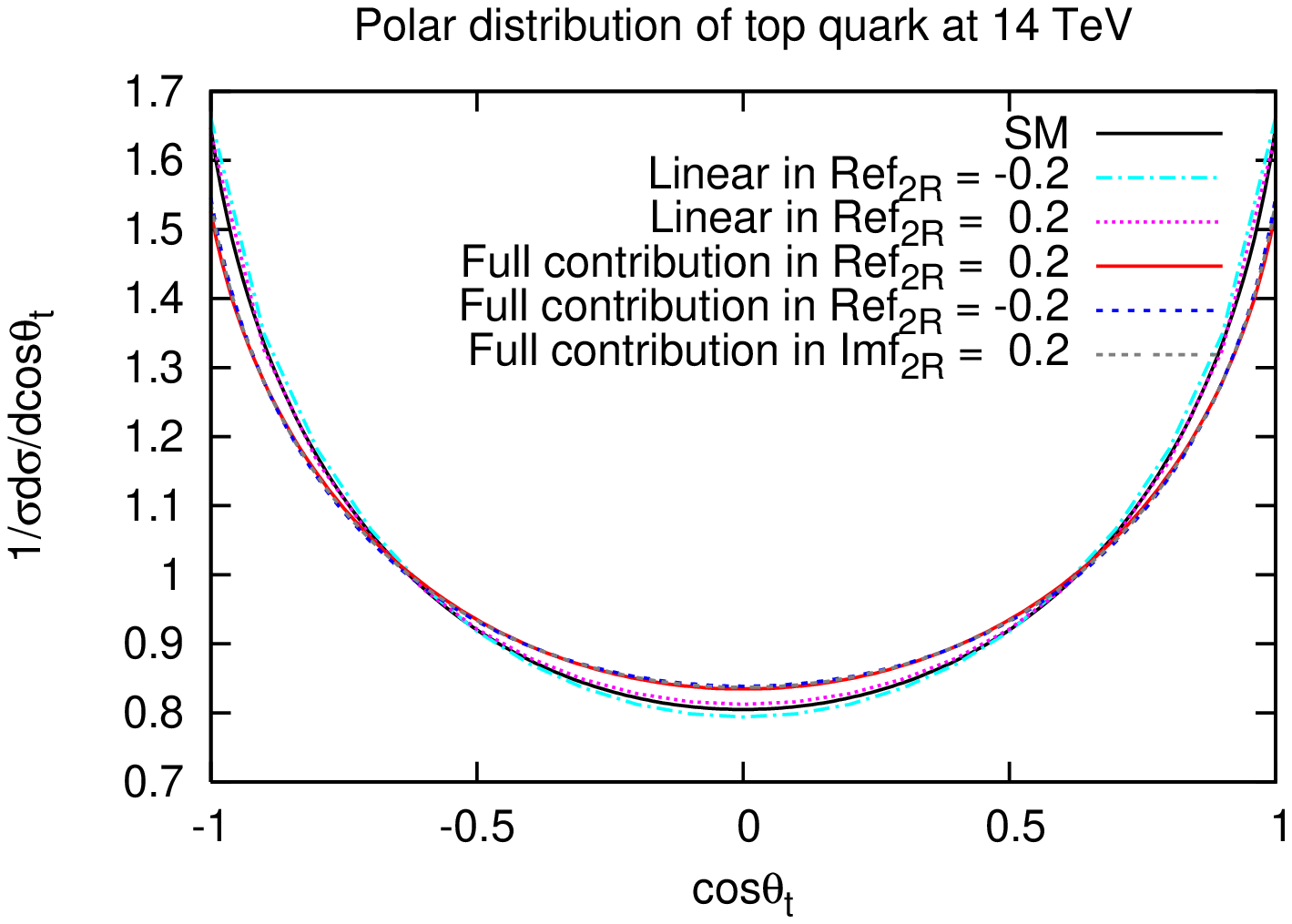} 
\caption{ The top-polar angular distributions for $tW^-$ production at the LHC
for two different cm energies, 7 TeV (left) and 14 TeV (right), for
different anomalous $tbW$ couplings.} 
\label{dist-top}
\end{center}
\end{figure}
The normalized polar distribution is plotted in Fig.
\ref{dist-top} for cm energies  7 TeV and 14 TeV.

As can be seen from Fig. \ref{dist-top}, the curves for the polar
distributions for the SM and for the anomalous couplings of magnitude
$0.2$ are separated from 
each other. The top distribution has no forward-backward asymmetry, the
colliding beams being identical. However, we can define an asymmetry utilizing the polar distributions of the top quark as 
\beq
\mathcal A_{\theta}^t=\frac{\sigma(|z|>0.5)-\sigma(|z|<0.5)}{\sigma(|z|>0.5)+\sigma(|z|<0.5)}
\eeq
\begin{figure}[htb]
\begin{center}
\includegraphics[angle=270,width=3.2in]{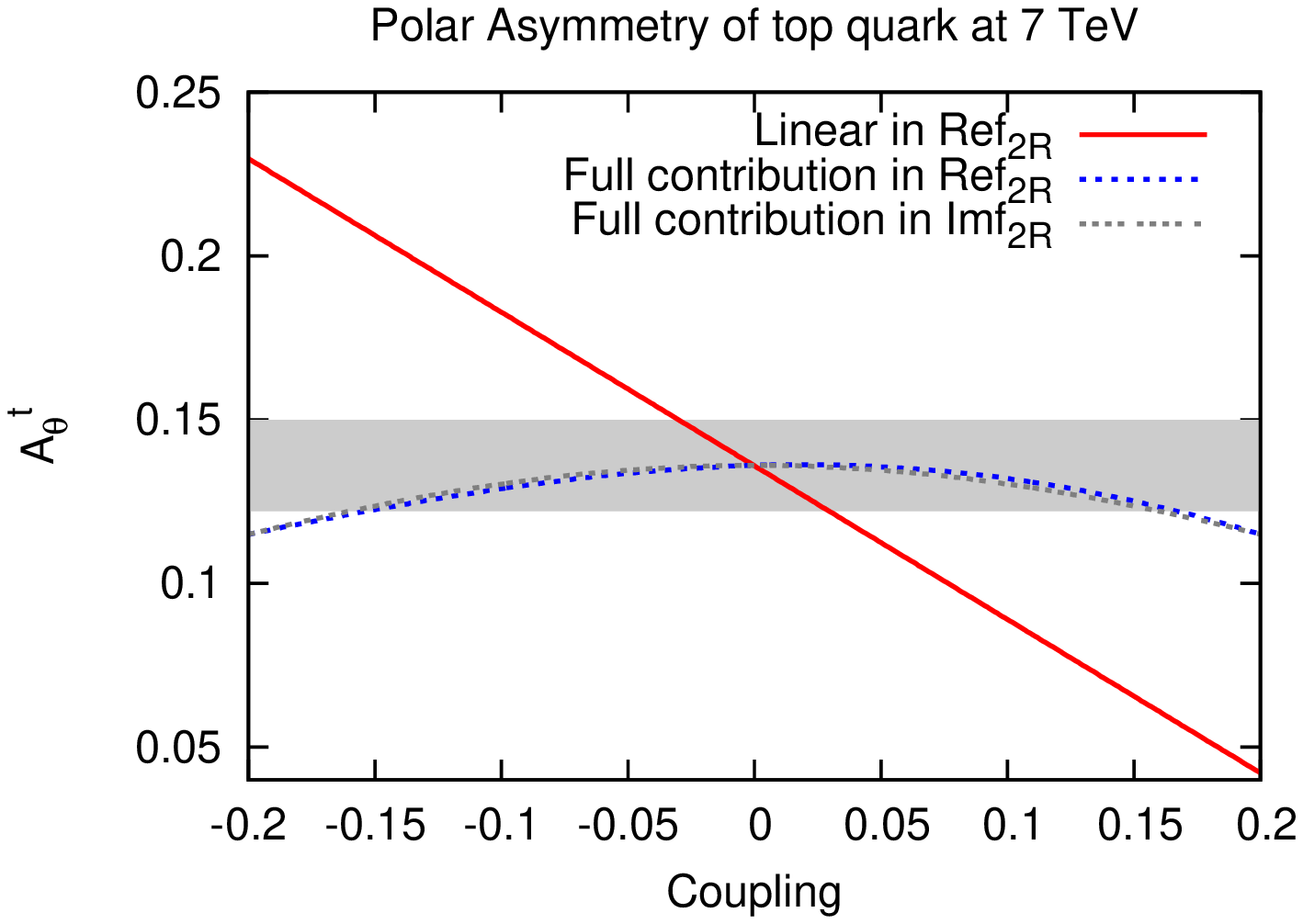} 
\includegraphics[angle=270,width=3.2in]{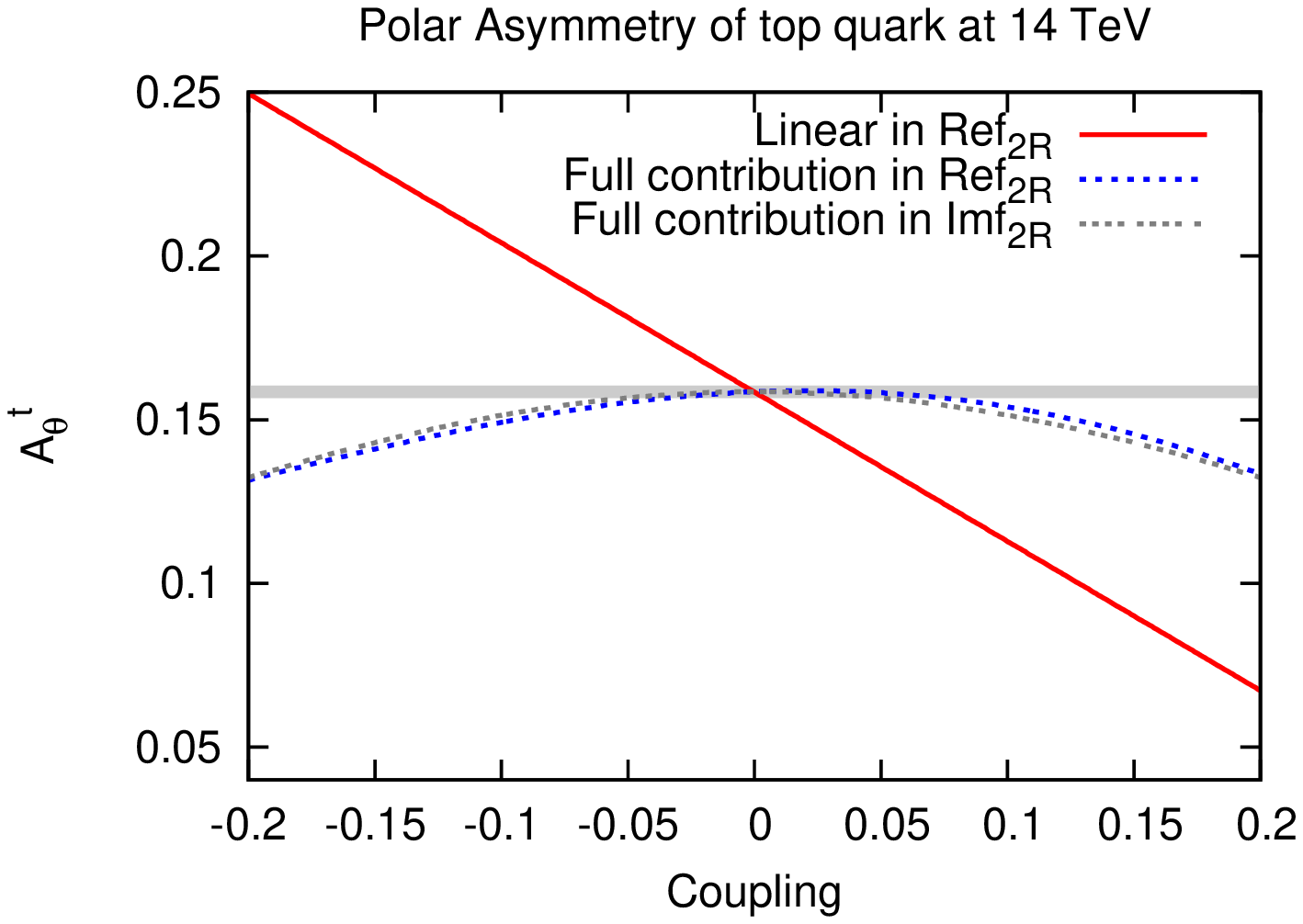} 
\caption{ The top polar asymmetries for $tW^-$ production at the LHC
for two different cm energies, 7 TeV (left) and 14 TeV (right), as a function of anomalous $tbW$ couplings.
The grey band corresponds to the top-polar asymmetry
predicted in the SM with a 1 $\sigma$ error interval. } 
\label{asym-top}
\end{center}
\end{figure}where $z$ is $\cos\theta_t$. We plot this asymmetry as a function of anomalous $tbW$ couplings for two cm of energies $\sqrt{s}=$ 
7 TeV and 14 TeV in Fig. \ref{asym-top}. $A_t^\theta$ deviates more than 1$\sigma$ from its SM value for Re$\rm f_{2R}$ and Im$\rm f_{2R}$ 
close to $\pm 0.2$ for $\sqrt{s}=$ 7 TeV, but is much more sensitive for $\sqrt{s}=$14 TeV.

The asymmetry $\mathcal A_{\theta}^t$ requires accurate determination of the top direction in
the lab frame and a quantitative estimate of its sensitivity to anomalous
couplings needs details of the efficiency of reconstruction of the direction. 
We do not study this asymmetry any further, but proceed to a discussion of top polarization. 

\subsection{Top polarization}

\begin{figure}[htb]
\begin{center}
\includegraphics[angle=270,width=3.2in]{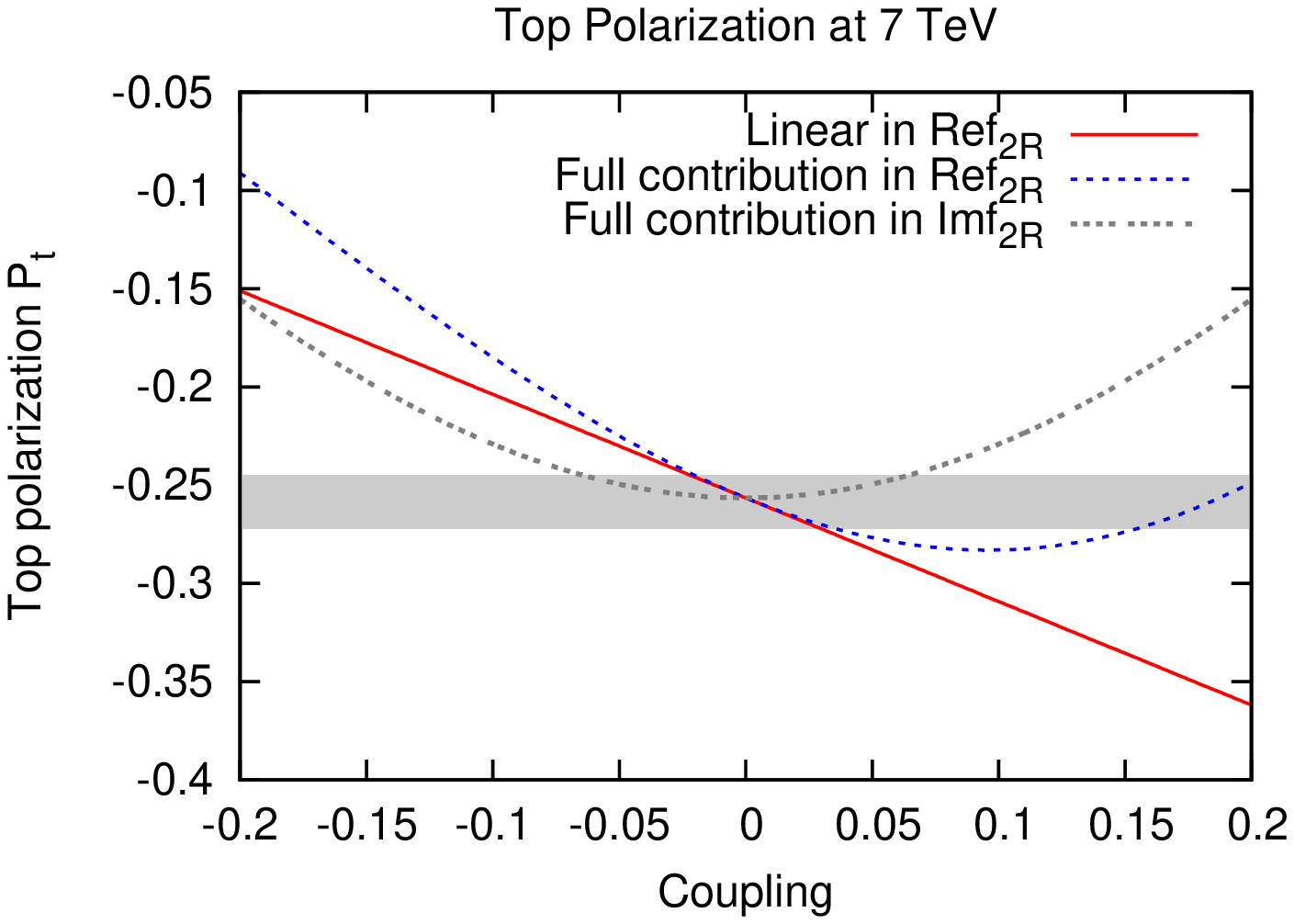} 
\includegraphics[angle=270,width=3.2in]{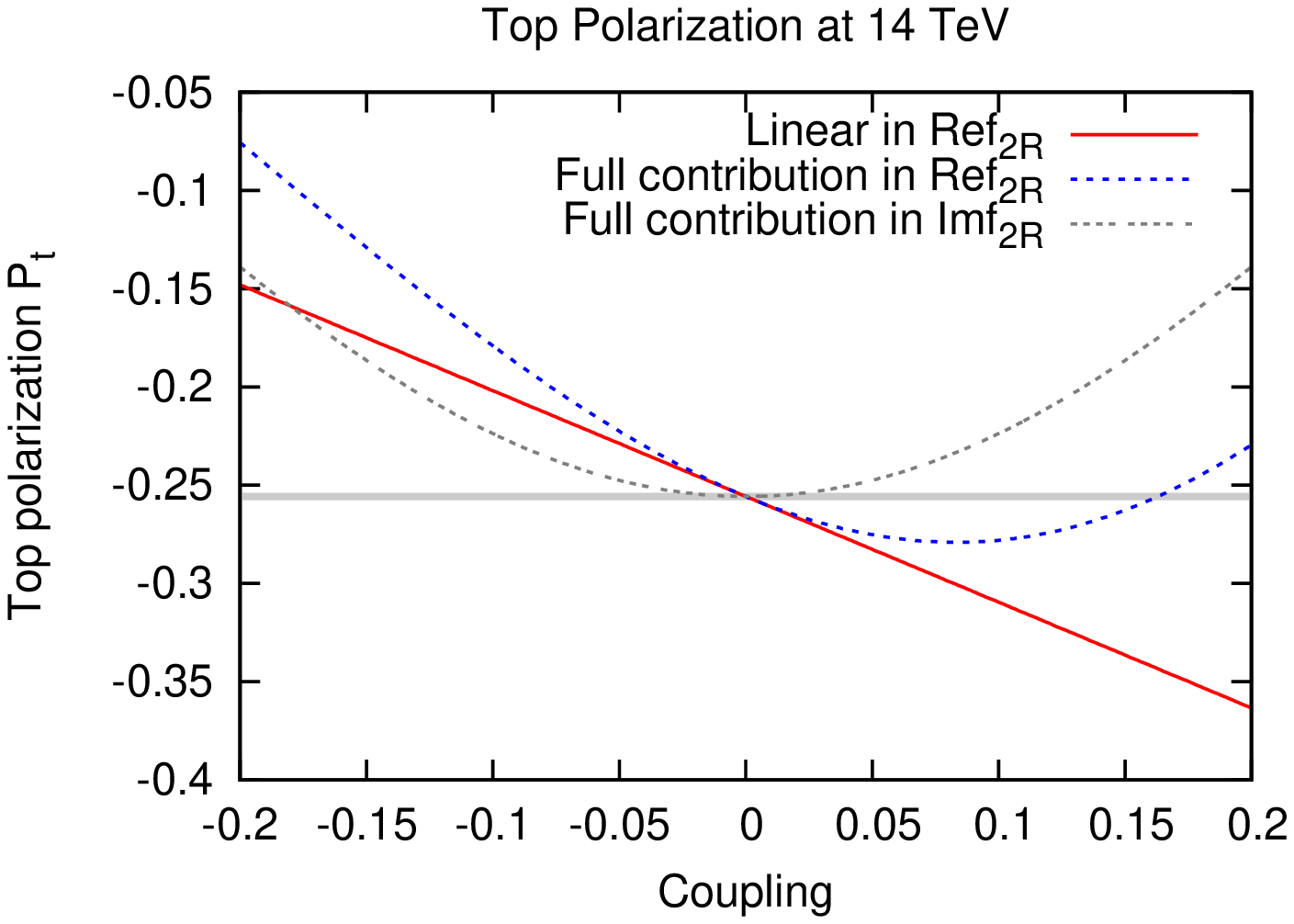} 
\caption{ The top polarization asymmetry for $tW^-$ production at the LHC
for two different cm energies, 7 TeV (left) and 14 TeV (right), as a function of anomalous $tbW$ couplings. 
The grey band corresponds to the top-polarization asymmetry predicted in the SM with a 1 $\sigma$ error interval. } 
\label{polcoup}
\end{center}
\end{figure}

The degree of longitudinal polarization $P_t$ of the top quark is given by
\begin{equation}
P_t=\frac{\sigma(+,+)-\sigma(-,-)}{\sigma(+,+)+\sigma(-,-)}.
\label{eta3def}
\end{equation}
This polarization asymmetry is shown in Fig. \ref{polcoup} as a function of 
anomalous couplings in the linear approximation, as well as without approximation, for $\sqrt{s}=$ 7 TeV and 14 TeV. 
As compared to the SM value of $-0.256$ for $\sqrt{s}=14$ TeV, 
the degree of longitudinal top polarization varies from $-0.07$ to
$-0.28$ for $\mathrm{Re}\frb$ varied over the range $-0.2$ to $+0.2$, while  
it varies from $-0.14$ to $-0.25$ 
for the same range of Im$\rm f_{2R}$, and  is symmetric about Im$\rm
f_{2R}=0$.

We notice that just as for the total cross section, $P_t$ is more sensitive to negative values of 
$\mathrm {Re}\frb$. 
Also, $P_t$ is 
equally sensitive to negative and positive values of $\mathrm{Im}\frb$. Thus $P_t$ can be a very good probe 
of $\mathrm {Re}\frb$ and $\mathrm {Im}\frb$ if it can be measured at the LHC. However, the standard measurement of $P_t$ 
requires reconstruction of the top rest frame which is a difficult task, and would entail reduction in efficiency. 
We will therefore investigate lab frame decay distributions for the measurement of anomalous couplings. 

All the quantities considered so far, viz., the total cross section, 
the top polar distribution and the top polarization, can only be measured
using information from the decay of the top. Both the polar
distribution and the top polarization would play a role in determining
the distributions of the decay products.
Our main aim 
is to devise observables which can be measured in the lab frame and give a good estimate of top polarization and 
thence probe anomalous $tbW$ couplings in single-top production. We proceed to construct such observables 
from kinematic variables of the charged lepton and $b$ quark produced
in the decay of the top.

\section{Charged-lepton distributions }
Top polarization can be determined by the angular distribution of its decay products. In the SM, the dominant decay mode is $t\to b W^+$, 
with a branching ratio (BR) of 0.998, with the $W^+$ subsequently decaying to $\ell^+ \nu_\ell$ (semileptonic decay, BR 1/9 for each lepton) or 
$u \bar{d}$, $c\bar{s}$ (hadronic decay, BR 2/3). The angular distribution of a decay product $f$ for a top-quark ensemble has the form 
 \begin{equation}
 \frac{1}{\Gamma_f}\frac{\textrm{d}\Gamma_f}{\textrm{d} \cos \theta _f}=\frac{1}{2}(1+\kappa _f P_t \cos \theta _f).
 \label{topdecaywidth}
 \end{equation}
 Here $\theta_f$ is the angle between fermion $f$ and the top spin vector in the top rest frame and $P_t$ (defined in Eq. \ref{eta3def}) 
is the degree of polarization of the top-quark ensemble. $\Gamma_f$ is the partial decay width and $\kappa_f$ is the spin analyzing power of $f$. 
Obviously, a larger $\kappa_f$ makes $f$ a more sensitive probe of the top spin. The charged lepton and the $d$ quark are the best spin analyzers 
with $\kappa_{\ell^+}=\kappa_{\bar{d}}=1$, while $\kappa_{\nu_\ell}=\kappa_{u}=-0.30$ and $\kappa_{b}=-\kappa_{W^+}=-0.39$, all $\kappa$ 
values being at tree level \cite{Bernreuther:2008ju}. Thus the $\ell^+$ or $d$ have the largest probability of being emitted in the direction of the top spin and the 
least probability in the direction opposite to the spin. Since at the LHC, lepton energy and momentum can be measured with high precision, 
we focus on leptonic decays of the top.

To preserve spin coherence while combining top production with decay, we make use of the spin density matrix formalism. Since $\Gamma_t/m_t \sim 0.008$, we 
can use the narrow width approximation (NWA) to write the cross section
in terms of the product of the $2\to 2$ production density matrix $\rho$
and the decay density matrix $\Gamma$ of the top as 
\begin{eqnarray}\label{NWA}
\overline{|{\cal M}|^2} = \frac{\pi \delta(p_t^2-m_t^2)}{\Gamma_t m_t}
\sum_{\lambda,\lambda'} \rho(\lambda,\lambda')\Gamma(\lambda,\lambda').
\end{eqnarray}
where and $\lambda,\lambda' =\pm 1$ denote (the sign of) the top helicity. 

The top production spin density matrix $\rho(\lambda,\lambda^{\prime})$ for process $gb\rightarrow tW^-$ is given in Appendix. 
We obtain the top decay density matrix $\Gamma(\lambda,\lambda')$ for the process $t\to b W^{+}\to b \ell^{+} \nu_\ell$ including 
anomalous $tbW$ couplings and write it in a Lorentz invariant form. We find 
\bea
\Gamma(\pm,\pm)&=& \nonumber g^4 \ |\Delta(p_W^2)|^2 [m_t^2-2(p_t\cdot p_\ell)]
 \Big[|\fla|^2\Big\{(p_t \cdot p_\ell) \mp m_t (p_\ell \cdot n_3)\Big\}\\\nonumber
&+&\mathrm{Re}\fla\mathrm{f_{2R}^*}\Big\{m_tm_W\mp\frac{m_t^2+m_W^2}{m_W}(p_\ell\cdot n_3)\mp\frac{2}{m_W}(p_b\cdot n_3)(p_t\cdot p_\ell)\Big\}\\
&+&\frac{|\frb|^2}{2}\Big\{m_W^2+\frac{m_t^2-2p_t\cdot p_\ell}{2}\mp2\left[(p_\ell\cdot n_3)+(p_b\cdot n_3)\right]\Big\}\Big],
\label{tdecaydia}
\eea
for the diagonal elements and
\bea
\Gamma(\mp,\pm)&=& \nonumber g^4 \ |\Delta(p_W^2)|^2 [m_t^2-2(p_t\cdot p_\ell)]
 \Big[|\fla|^2\left\{- m_t [p_\ell \cdot (n_1\mp i n_2)]\right\}\\\nonumber
&-&\mathrm{Re}\fla\mathrm{f_{2R}^*}\Big\{\frac{m_t^2+m_W^2}{m_W}[p_\ell \cdot (n_1\mp i n_2)]-\frac{2}{m_W}[p_b \cdot (n_1\mp i n_2)](p_t\cdot p_\ell)\Big\}\\
&-&|\frb|^2\Big\{\left[(p_\ell+p_b) \cdot (n_1\mp i n_2)\right]\Big\}\Big],
\label{tdecayoffdia}
\eea
for the off-diagonal ones. Here $\Delta(p_W^2)$ is the $W$ boson
propagator, for which we will use the narrow-width approximations for
our numerical calculations, and $n^{\mu}_{i}$'s ($i=1,2,3$) are  
spin four-vectors for the top with four-momentum $p_t$, corresponding to
rest-frame spin quantization axes $x$, $y$ and $z$ respectively,
with the properties 
$n_i \cdot n_j =-\delta_{ij}$ and $n_i \cdot p_t =0$. 
In the top rest frame they take the standard form $n^{0}_{i}=0$, 
$n^k_i = \delta_{i}^{k}$.

The full details of the factorization of the differential cross section 
for top production followed by its decay in a generic production process 
into production and decay parts is given in Ref. \cite{Godbole:2006tq}. 
Here we use the same formalism for single-top production and its decay and 
write the partial cross section in the parton cm of frame as 
\bea
d\sigma&=&\frac{1}{32(2\pi)^4 \Gamma_t m_t} 
\int \left[ \sum_{\lambda,\lambda'}
\frac{d\sigma(\lambda,\lambda')}{d\cos\theta_t} \
\left(\frac{\langle\Gamma(\lambda,\lambda')\rangle}{p_t \cdot p_\ell}\right) \right] \ 
d\cos\theta_t \ d\cos\theta_\ell \ d\phi_\ell\ E_\ell dE_\ell \ dp_W^2,
\label{dsigell}
\eea
where the $b$-quark energy integral is replaced by an integral 
over the invariant mass $p_W^2$ of the $W$ boson, its polar-angle
integral is carried out using the Dirac delta function of Eq.
\ref{NWA}, and the average over its azimuthal angle is denoted by the
angular brackets.
$d\sigma(\lambda,\lambda')/d\cos\theta_t$ is proportional to
$\rho(\lambda, \lambda')$, with the normalization chosen so that
$d\sigma(\lambda,\lambda)/d\cos\theta_t$ is the differential 
cross section for the $2\to2$ process of $tW^-$ production with helicity
index $\lambda$ of the top. 

\subsection{Angular distributions of charged leptons }

We evaluate top decay in the rest frame of the top quark with the
$z$ axis as the spin quantization axis, which would also be the
direction of the boost required to go to the parton cm frame. 
In the rest frame of the top quark, the diagonal and off-diagonal elements of decay density matrix, after averaging over 
the azimuthal angle of $b$ quark w.r.t. the plane of top and lepton momenta, are given by 
\bea\label{decay-diag}
\langle\Gamma(\pm,\pm)\rangle&=&\nonumber g^4m_tE_\ell^0|\Delta_W(p_W^2)|^2(m_t^2-2p_t\cdot p_\ell)\Bigg[
\left\{|\fla|^2+\mathrm{Re}\fla\mathrm{f_{2R}^*}\frac{m_t \ m_W}{p_t\cdot p_\ell}\right\}(1\pm\cos\theta^0_{\ell})\\
&-&|\frb|^2\left(1-\frac{m_t^2+m_W^2}{2p_t\cdot p_\ell}\right)(1\mp\cos\theta^0_{\ell})
\pm|\frb|^2\frac{m_W^2m_t^2}{2(p_t\cdot
p_\ell)^2}\cos\theta^0_{\ell}\Bigg],\\
\langle\Gamma(\pm,\mp)\rangle&=& \nonumber g^4m_tE_\ell^0|\Delta_W(p_W^2)|^2(m_t^2-2p_t\cdot p_\ell)\sin\theta^0_{\ell}e^{\pm i\phi^0_{\ell}}
\Bigg[|\fla|^2+\mathrm{Re}\fla\mathrm{f_{2R}^*}\frac{m_t \ m_W}{p_t\cdot p_\ell}\\\label{decay-offdiag}
&+&|\frb|^2\left\{1-\frac{m_t^2+m_W^2}{2p_t\cdot p_\ell}+\frac{m_W^2m_t^2}{2(p_t\cdot p_\ell)^2}\right\}
\Bigg],
\eea
where averaging over the azimuthal angle of the $b$ quark w.r.t. the
plane of top and lepton momenta, denoted by angular brackets, is most
conveniently carried out in a coordinate system defined with the $z$ axis along the lepton momentum direction. 

In the limit of small anomalous coupling $\frb$, we see from Eqs.
\ref{decay-diag} and \ref{decay-offdiag} that if we drop quadratic terms
in $\frb$, 
$\langle\Gamma(\lambda,\lambda^{\prime})\rangle$ factorizes into a pure
angular part $\mathcal A(\lambda,\lambda^{\prime})$, which depends on
helicities,  
and a lepton-energy dependent part 
which does not depend on the helicities, where
 \bea\label{fel}
\mathcal A(\pm,\pm)&=&(1\pm\cos\theta^0_\ell),~~~~
\mathcal A(\pm,\mp)=\sin\theta^0_\ell e^{\pm i\phi^0_\ell}.
 \eea

The factorization of $\langle\Gamma(\lambda,\lambda^{\prime})\rangle$ into $\mathcal A(\lambda,\lambda^{\prime})$ 
and a helicity-independent part in the rest frame of the top quark
implies that since the corrections from anomalous couplings reside in
the helicity-independent part, they are identical to those of the total
width appearing in the denominator of the angular distribution, and
cancel. This leads to the result of \cite{decoupling,Godbole:2006tq} that the energy averaged 
lepton angular distributions are insensitive to the new physics in top-quark decay in any top production process. 

We study the angular distribution of the charged lepton in the lab frame both in the linear approximation of the anomalous couplings 
as well as with full contributions of the anomalous couplings without approximation. 

We first obtain the angular distribution of the charged 
lepton in the parton cm frame, by integrating over the lepton energy, with
limits given by $m_W^2<2(p_t\cdot p_\ell) < m_t^2$. This integral can be done analytically, giving the following expression for 
the differential cross section in the parton cm frame:
\bea\label{angdist}
\frac{d\sigma}{d\cos\theta_t \ d\cos\theta_\ell \ d\phi_\ell} &=&  \frac{1}{32 \ \Gamma_t m_t} \ \frac{1}
{(2\pi)^4}\int \left[ \sum_{\lambda,\lambda'} \frac{d\sigma
(\lambda,\lambda')}{d \cos\theta_t} 
 g^4 \mathcal{A}^\prime(\lambda,\lambda') \right]
|\Delta(p_W^2)|^2 dp_W^2,
\eea
where 
\bea\label{angmat1}
 \mathcal A^\prime (\pm,\pm)&=&\nonumber\frac{m_t^6}{24(1-\beta_t
\cos\theta_{t\ell})^3 E_t^2}
\Big[(1-r^2)^2  
\big\{(1\pm\cos\theta_{t\ell})(1\mp\beta_t)\left[|\fla|^2(1+2r^2)+6r
\mathrm{Re\fla f_{2R}^*}\right] \\
&+&\nonumber |\frb|^2(2+r^2)(1\mp \cos\theta_{t\ell}) (1\pm\beta_t)\big\}\\
&\mp& 12r^2 
|\frb|^2\left(1-r^2+2 \log r\right)(\cos\theta_{t\ell} - \beta_t)\Big],\\
\label{angmat2}
\mathcal A^\prime (\pm,\mp)&=&\nonumber\frac{m_t^7}{24(1-\beta_t
\cos\theta_{t\ell})^3 E_t^3}\sin\theta_{t\ell} e^{\pm i\phi_\ell}
\Big[(1-r^2)^2\big\{|\fla|^2(1+2r^2)+6r \ \mathrm{Re\fla f_{2R}^*}\\
&-&|\frb|^2(2+r^2)\big\}- 12r^2 \ |\frb|^2\left(1-r^2+2 \log r\right)\Big].
 \eea
Here $r=m_W/m_t$ and $\cos\theta_{t\ell}$ is the angle between the top
quark and the charged lepton from the top decay in the parton cm frame, 
given by
\begin{equation}
 \cos \theta_{t \ell}=\cos \theta_t \cos \theta_{\ell}+\sin \theta_t \sin \theta_{\ell} \cos \phi_{\ell},
\label{costhetatl}
\end{equation}
where $\theta_\ell$ and $\phi_\ell$ are the lepton polar and azimuthal angles.

In the lab frame, the lepton polar angle is defined w.r.t. either of the
beam direction and the azimuthal angle is 
defined with respect to the top-production plane chosen as the $x$-$z$ plane, with beam direction as the $z$ axis and the 
convention that the $x$ component of the top momentum is positive. At the LHC, which is a symmetric collider, it is not possible to define
a positive direction of the $z$ axis. Hence lepton angular distribution is symmetric under interchange of
$\theta_{\ell}$ and $\pi-\theta_{\ell}$ as well as of $\phi_{\ell}$ and $2\pi-\phi_{\ell}$.
The lab frame expression for the differential cross section is obtained
from Eq. \ref{angdist} by an appropriate Lorentz transformation and
integration over the parton densities.
\begin{figure}[t]
\begin{center}
\includegraphics[angle=270,width=3.2in]{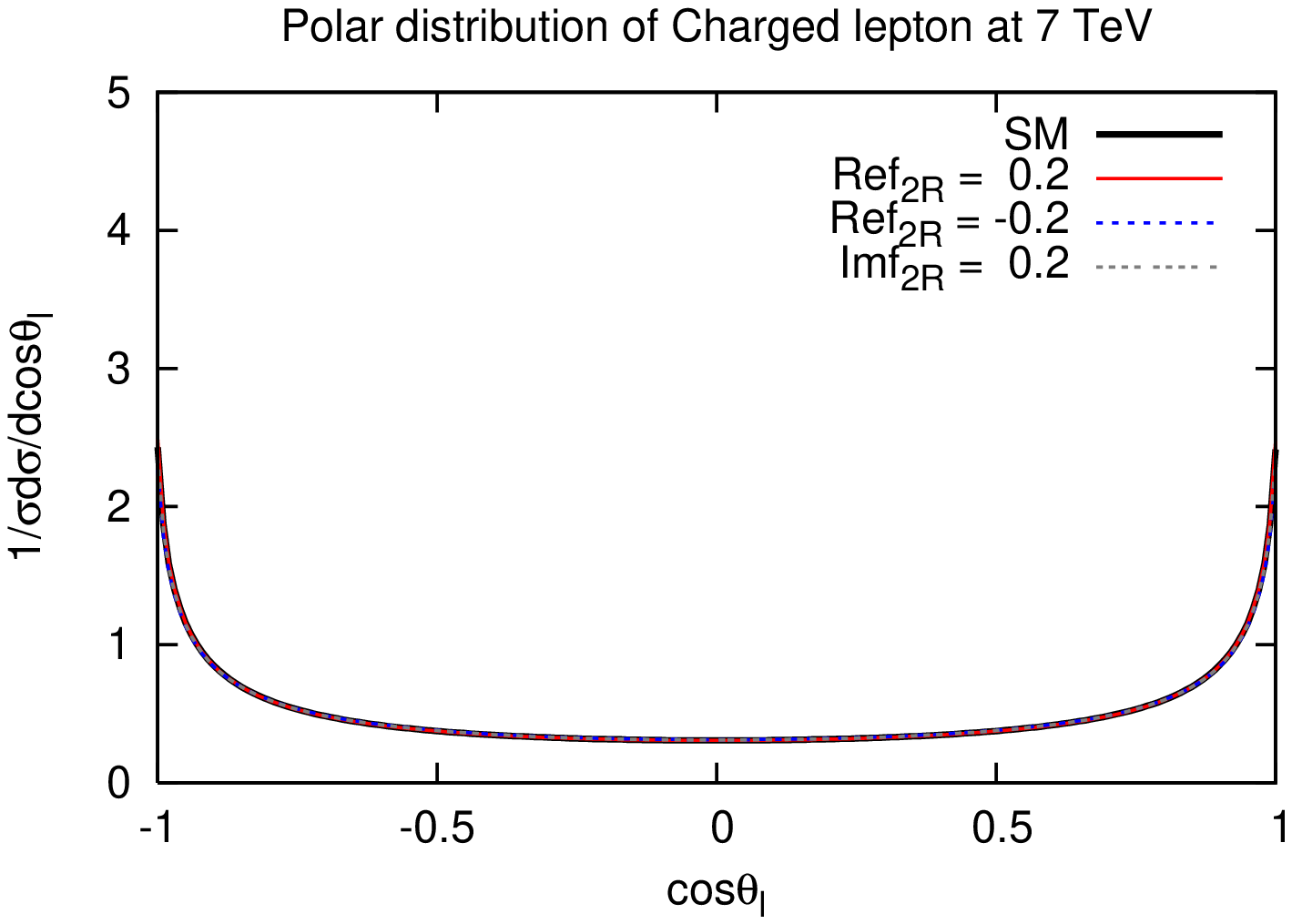} 
\includegraphics[angle=270,width=3.2in]{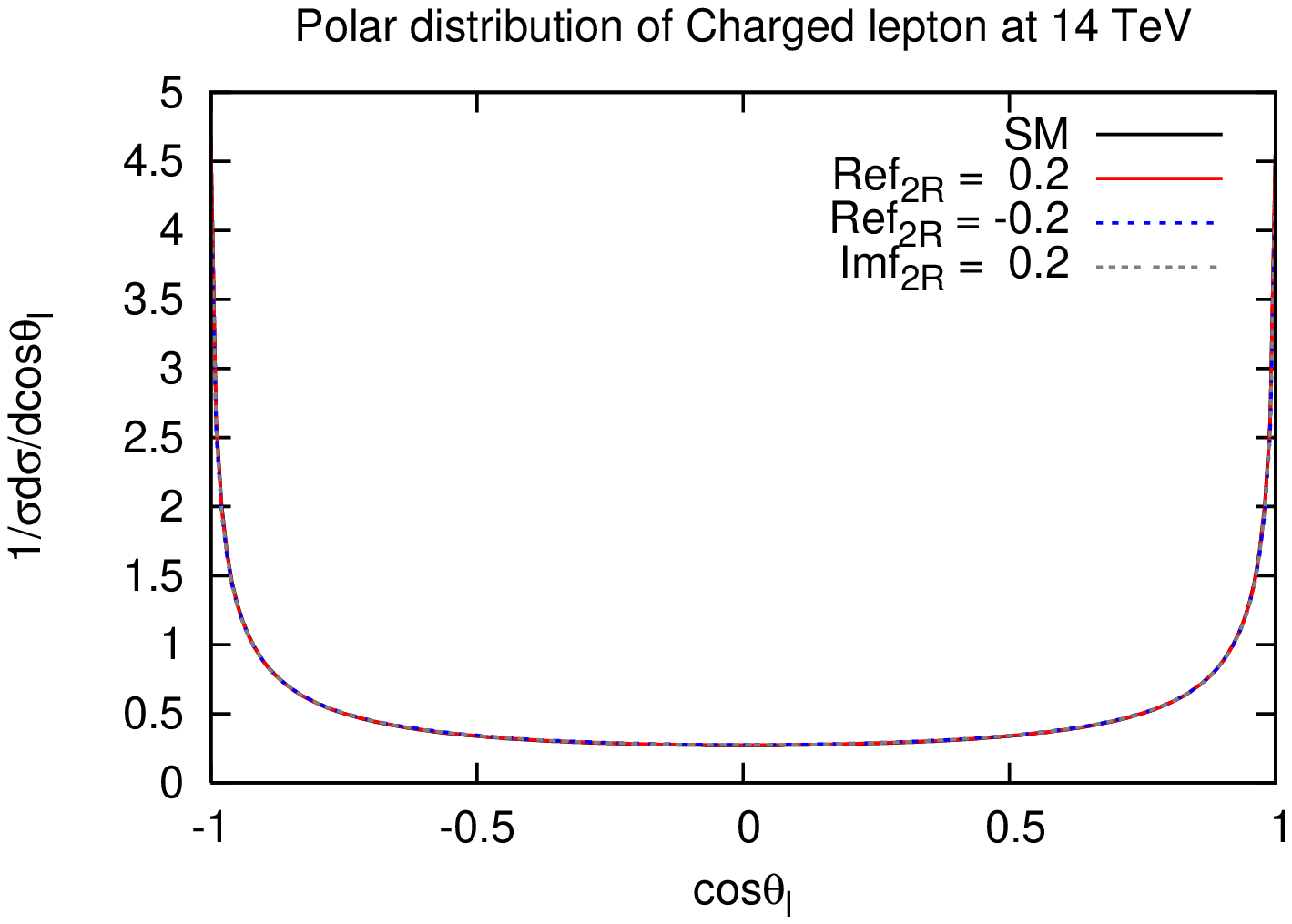} 
\caption{ The normalized polar distribution of the charged lepton in
$tW^-$  production at the LHC
for two different cm energies, 7 TeV (left) and 14 TeV (right), for
different anomalous $tbW$ couplings.} 
\label{dist-pol}
\end{center}
\end{figure}
\begin{figure}[h]
\begin{center}
\includegraphics[angle=270,width=3.2in]{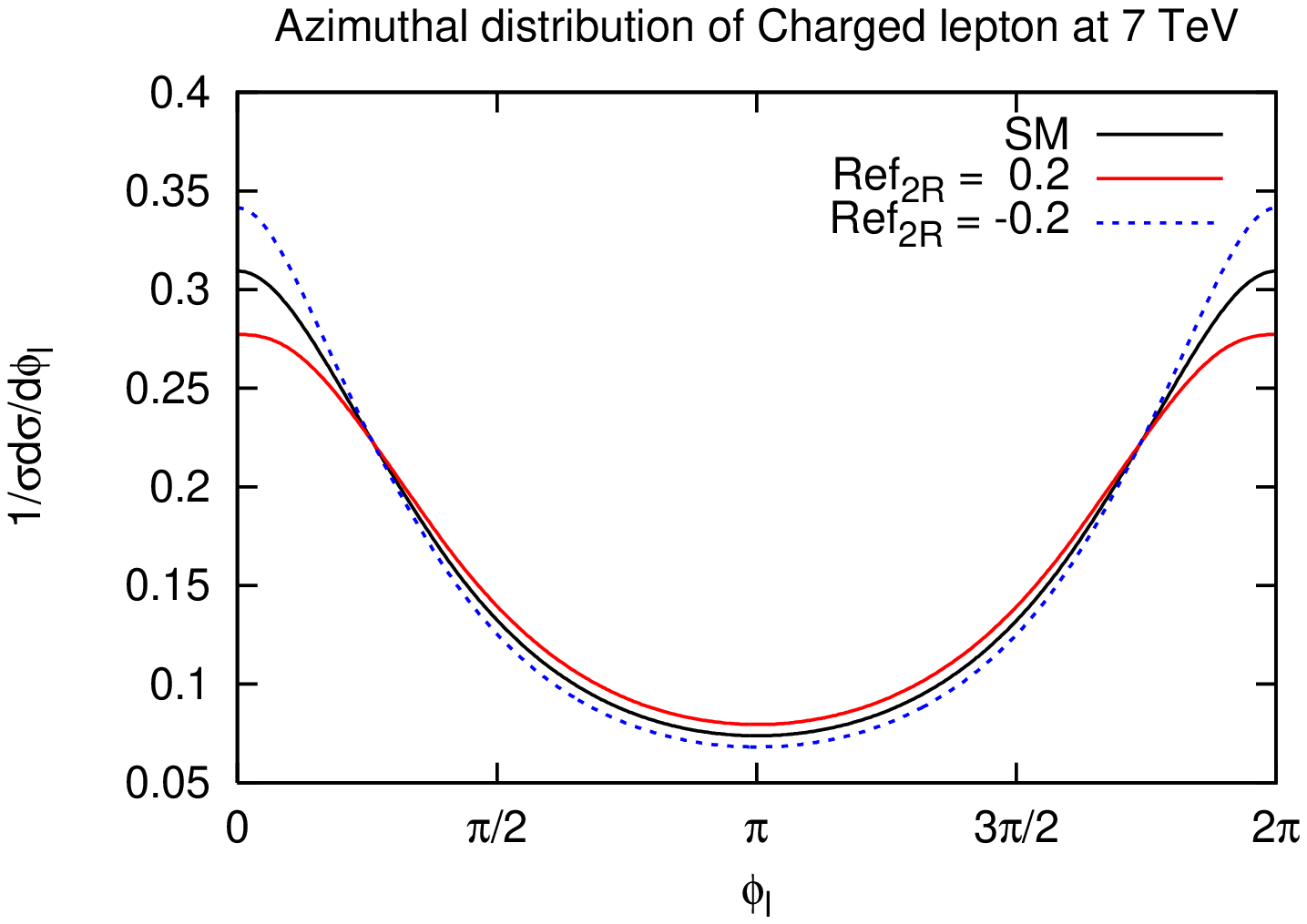} 
\includegraphics[angle=270,width=3.2in]{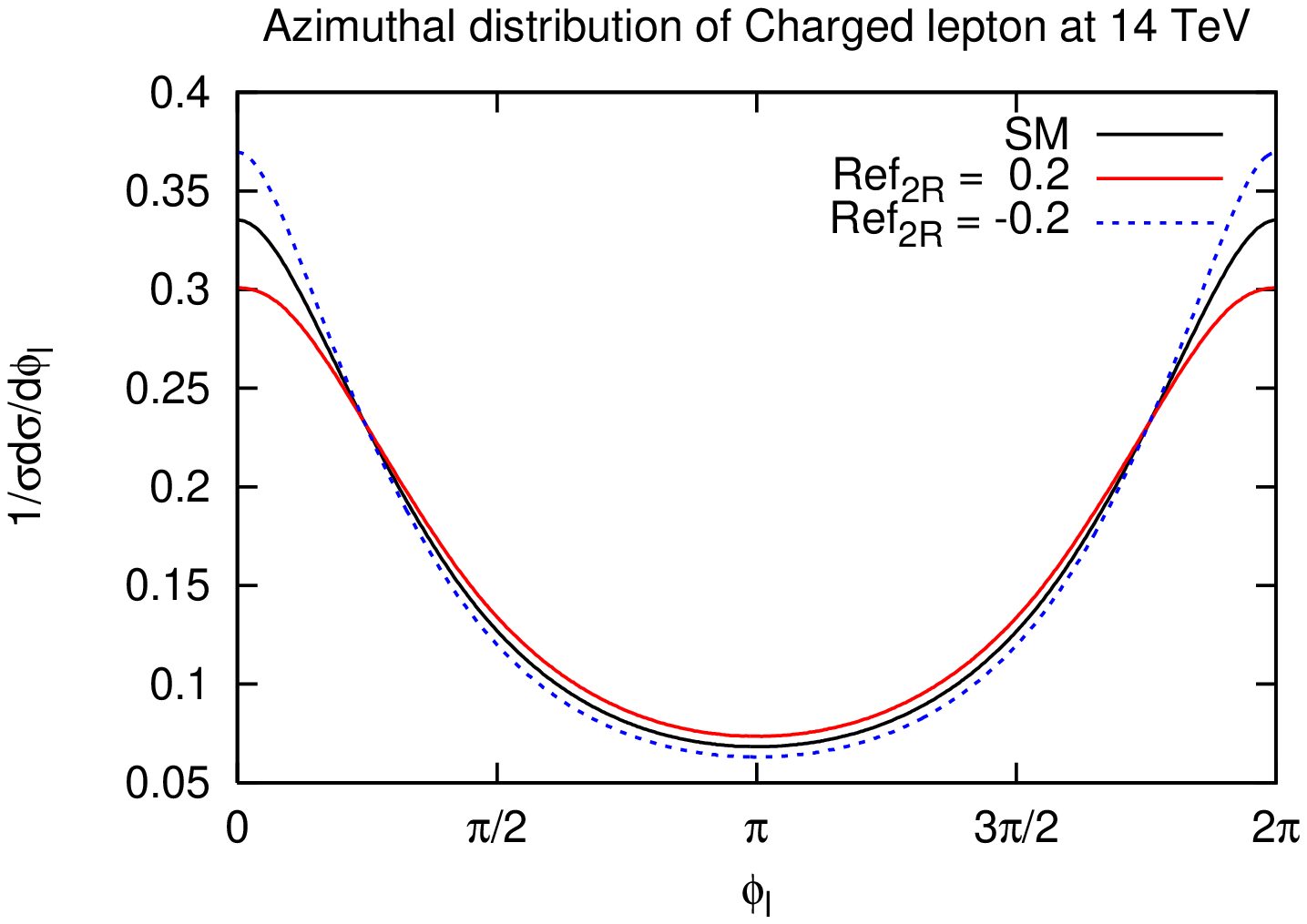} 
\caption{ The normalized azimuthal distribution of the charged lepton in
$tW^-$  production at the LHC for two different cm energies, 7 TeV (left) and 14 TeV (right), for
different  anomalous $tbW$ couplings in the linear approximation.
$\rm Im\frb$ does not contribute in the linear order.} 
\label{dist-azi-lin}
\end{center}
\end{figure}

We first look at the polar distribution of the charged lepton and  the effect on it of anomalous $tbW$ 
couplings. As can be seen from Fig. \ref{dist-pol}, where we plot the polar distribution 
for two cm of the LHC energies $\sqrt{s}=$ 7 TeV and 14 TeV, the normalized distributions are 
insensitive to anomalous $tbW$ couplings.  

\begin{figure}[htb]
\begin{center}
\includegraphics[angle=270,width=3.2in]{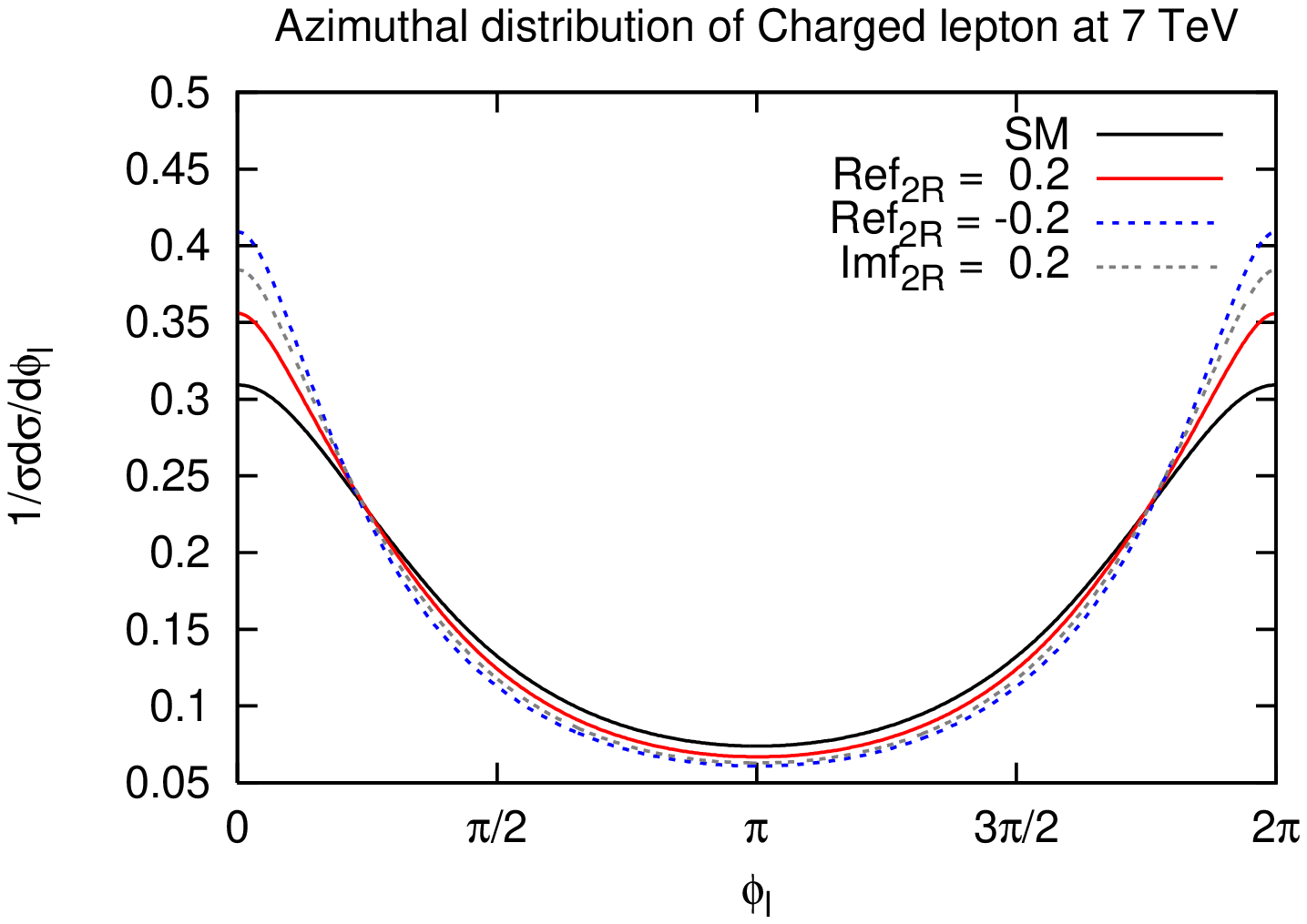} 
\includegraphics[angle=270,width=3.2in]{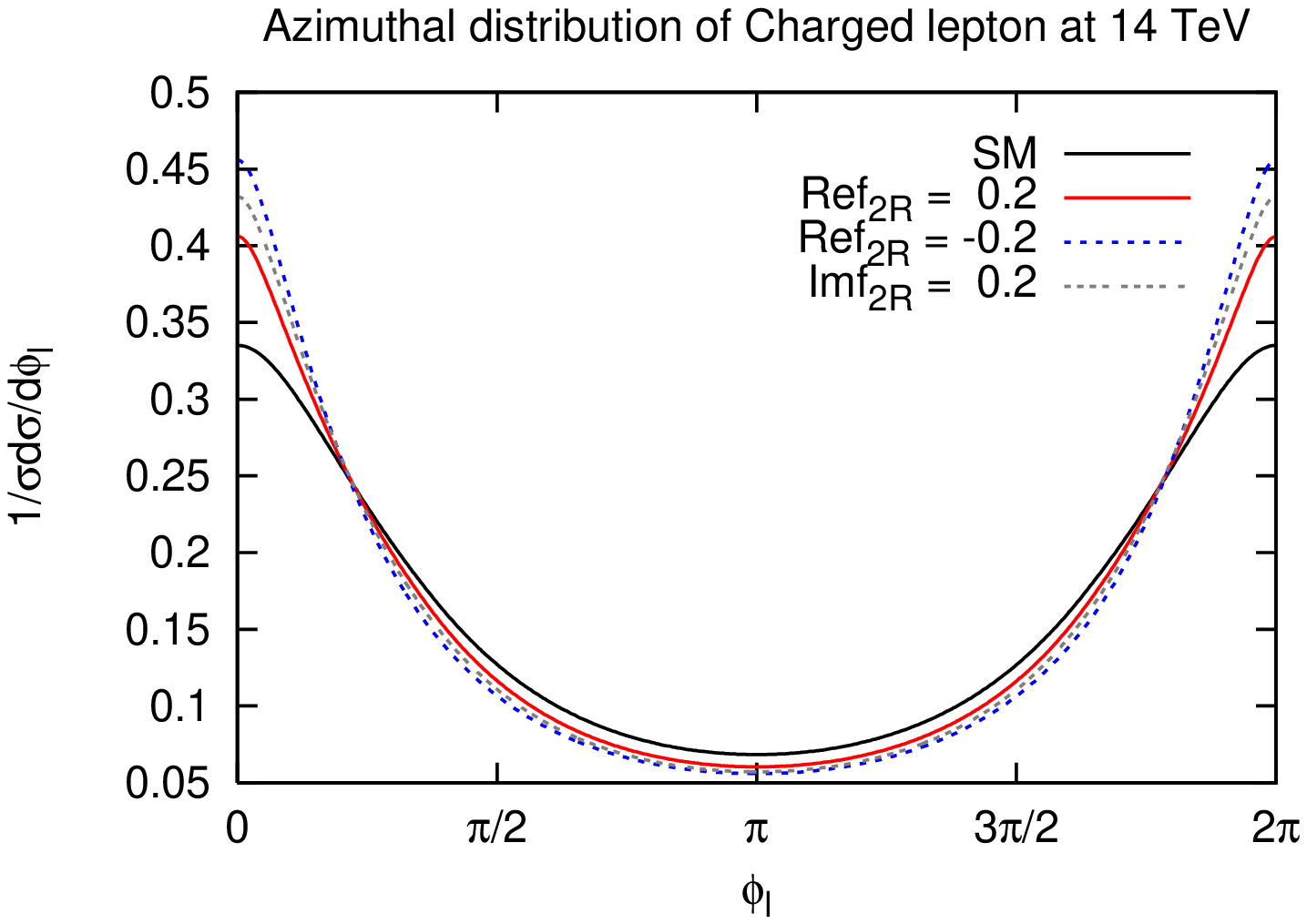} 
\caption{ The normalized azimuthal distribution of the charged lepton in
$tW^-$  production at the LHC for two different cm energies, 7 TeV (left) and 14 TeV (right), for
different  anomalous $tbW$ couplings without the linear approximation} 
\label{dist-azi}
\end{center}
\end{figure}

We next look at the contributions of anomalous couplings to 
the azimuthal distribution of the charged lepton. In Fig. \ref{dist-azi-lin}, we show the normalized azimuthal distribution of 
the charged lepton in a linear approximation of the couplings for $\sqrt{s} =$ 7 TeV and 14 TeV 
for different values of $\mathrm{Re}\frb$. At linear order, contributions of all other couplings vanish in the limit of vanishing bottom mass. 
In Fig. \ref{dist-azi}, we show the normalized azimuthal distribution of
the charged lepton including higher-order terms in the couplings for 
$\sqrt{s} =$ 7 TeV and 14 TeV for different values of $\mathrm{Re}\frb$ and $\mathrm{Im}\frb$. We see that the curves for real and 
imaginary parts of the anomalous coupling $\frb$ peak near
$\phi_{\ell}=0$ and $\phi_{\ell}=2\pi$.
The curves are well separated at the peaks for the chosen values of the anomalous $tbW$ couplings 
and are also well separated from the curve for the SM. 
We define an azimuthal asymmetry for the lepton to quantify 
these differences in the distributions by
\begin{equation}
 A_{\phi}=\frac{\sigma(\cos \phi_\ell >0)-\sigma(\cos
\phi_\ell<0)}{\sigma(\cos \phi_\ell >0)+\sigma(\cos \phi_\ell<0)},
\label{aziasy}
\end{equation}
\begin{figure}[h]
\begin{center}
\includegraphics[angle=270,width=3.2in]{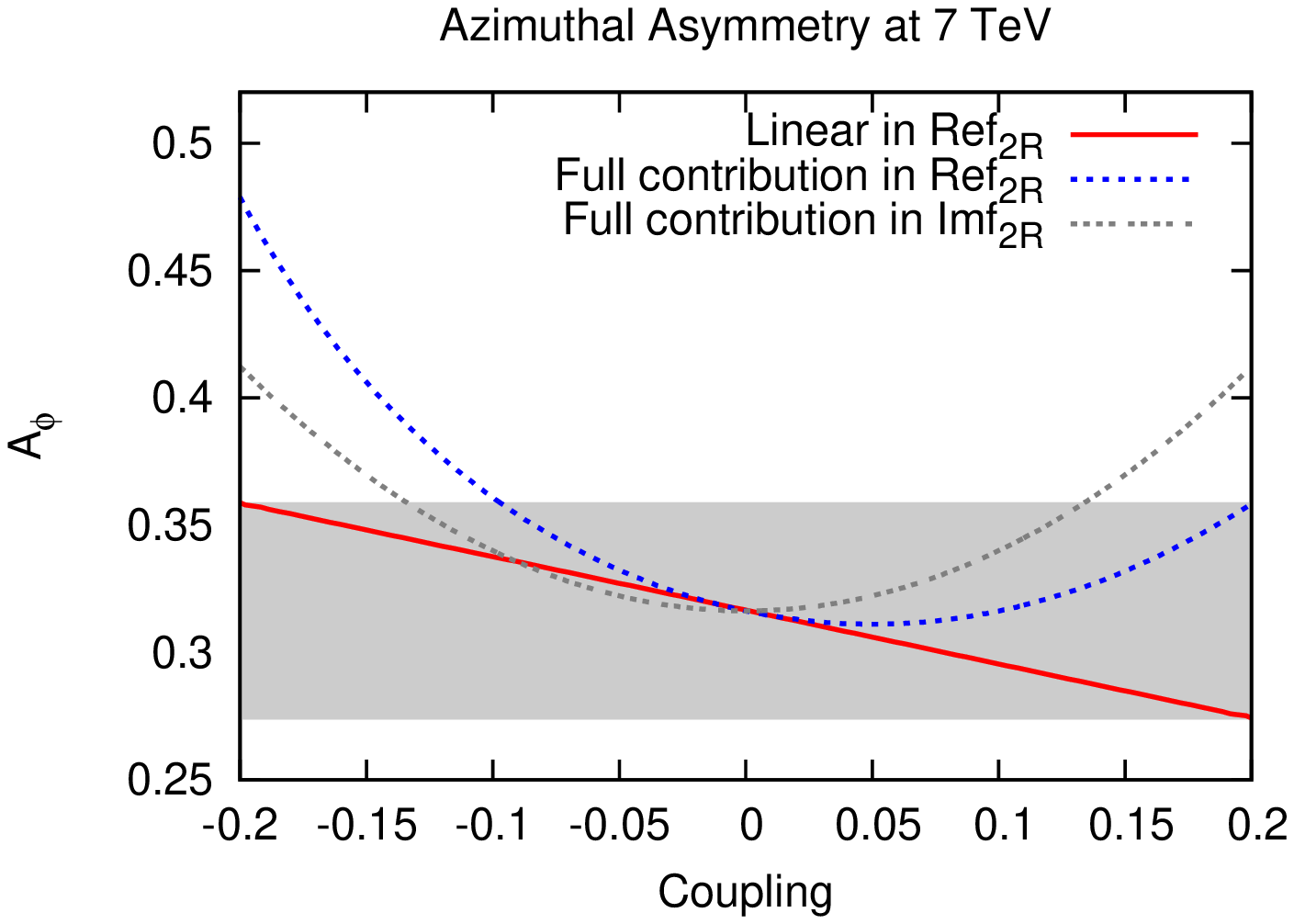} 
\includegraphics[angle=270,width=3.2in]{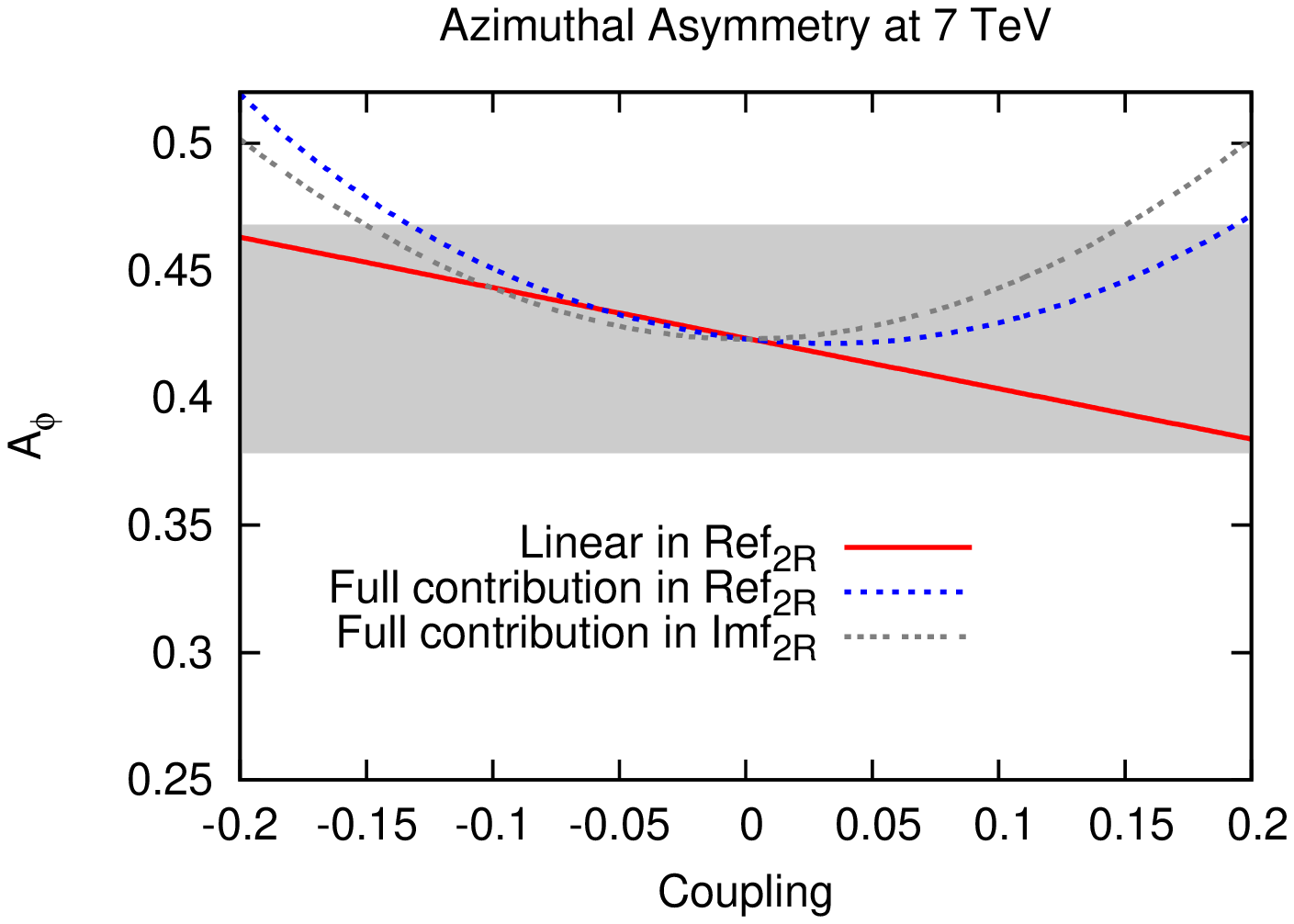} 
\caption{ The azimuthal asymmetries of the charged lepton in $tW^-$
production at the LHC
for cm energy 7 TeV without lepton cuts (left) and with cuts (right), as a function of anomalous $tbW$ couplings.
The grey band corresponds to the azimuthal asymmetry predicted in the SM with  1 $\sigma$ error interval. } 
\label{aziasy7lepton}
\end{center}
\end{figure}
where the denominator is the total cross section. Plots of $A_{\phi}$
as a function of the couplings
with and without cuts on the lepton momenta are shown in Fig. \ref{aziasy7lepton} 
for a cm energy of 7 TeV. In the former case, 
the rapidity and transverse momentum acceptance cuts on the decay 
lepton that we have used are $|\eta|<2.5,\,p_{T}^\ell>20$ GeV.
\begin{figure}[h]
\begin{center}
\includegraphics[angle=270,width=3.2in]{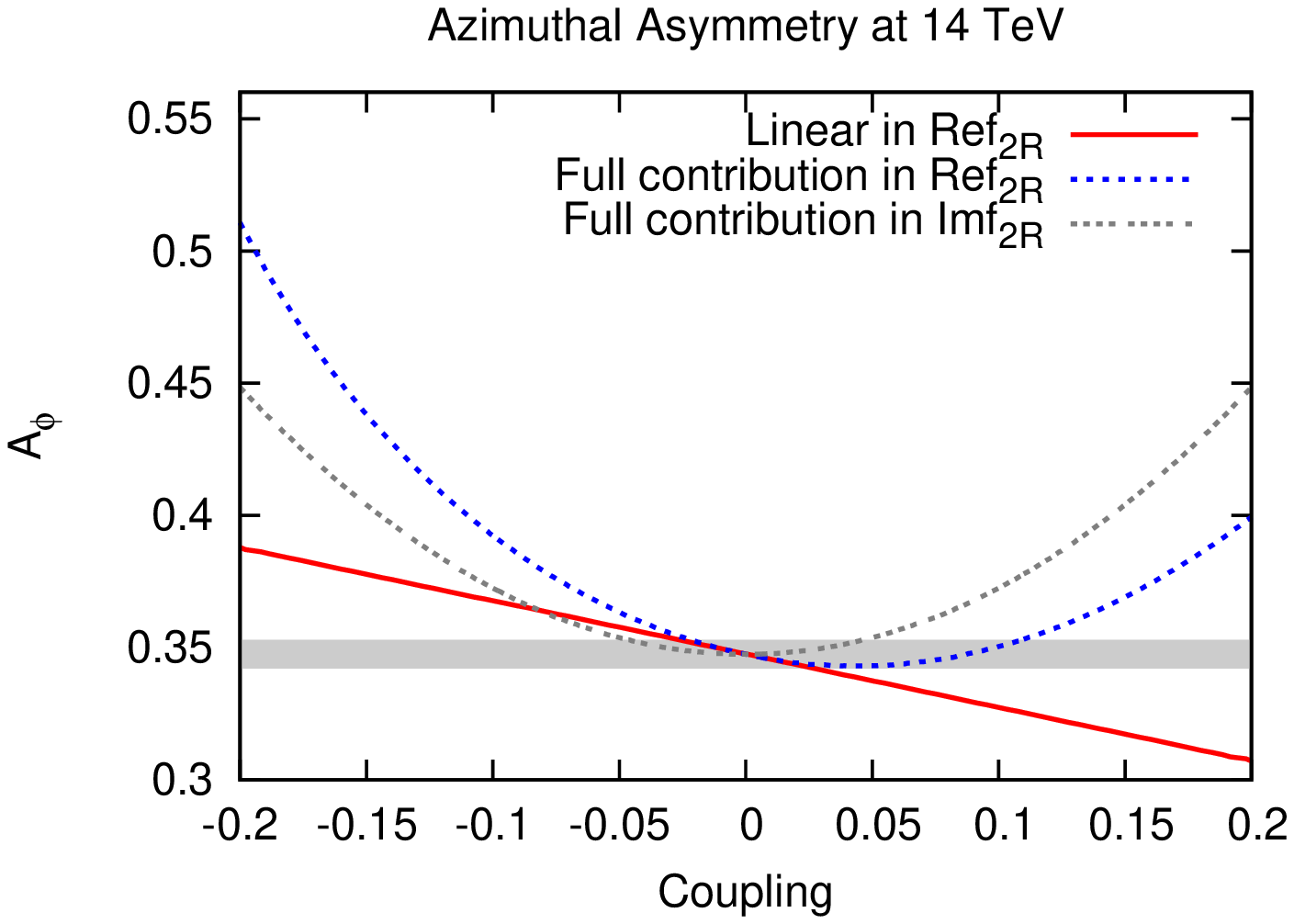} 
\includegraphics[angle=270,width=3.2in]{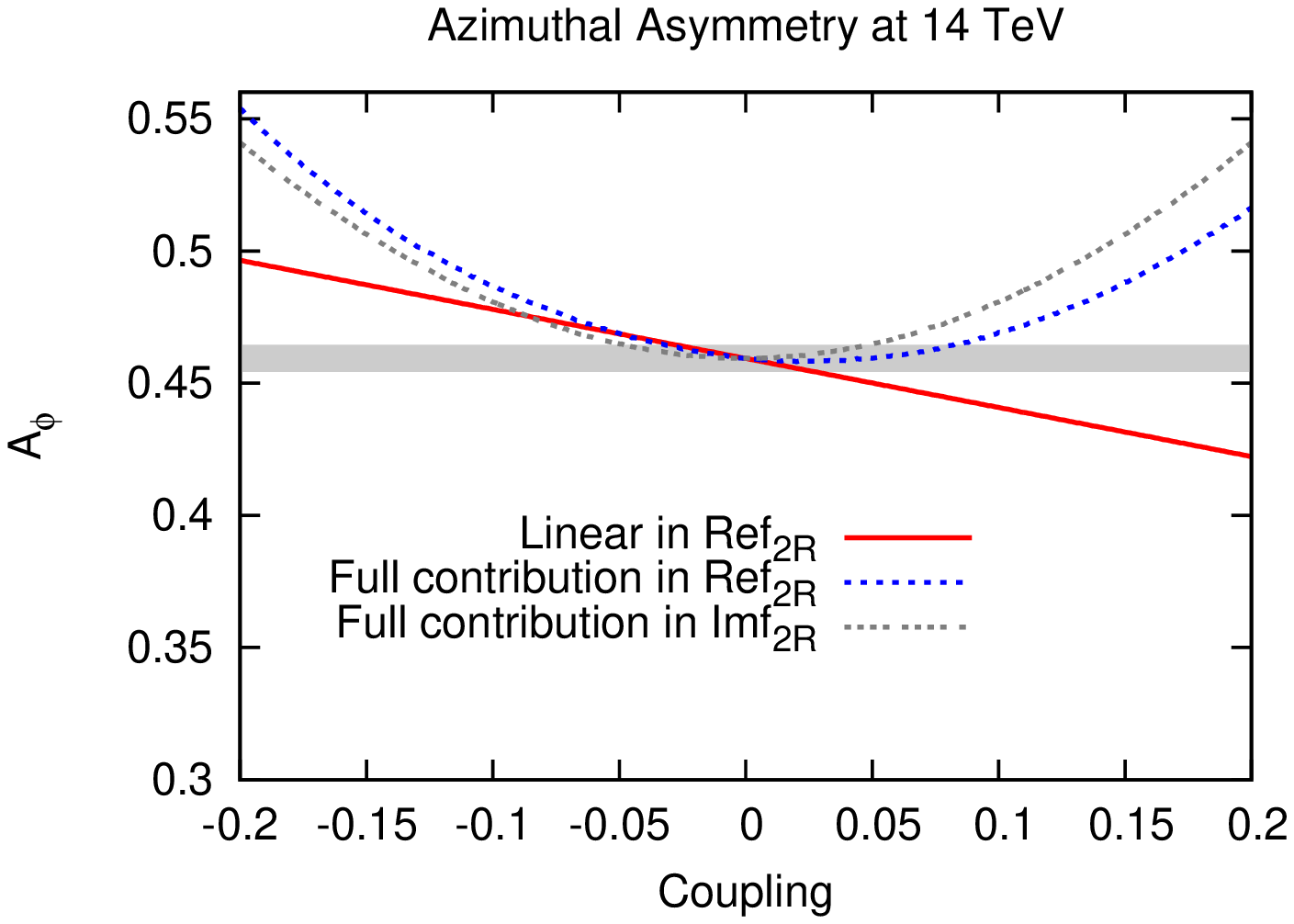} 
\caption{ The azimuthal asymmetries of the charged lepton in $tW^-$
production at the LHC
for cm energy 14 TeV without lepton cuts (left) and with cuts (right), as a function of anomalous $tbW$ couplings.
 The grey band corresponds to the azimuthal asymmetry
predicted in the SM with a 1$\sigma$ error interval. } 
\label{aziasy14lepton}
\end{center}
\end{figure}
The corresponding plots at $\sqrt{s} = $14 TeV with and without lepton cuts are shown 
in the Fig. \ref{aziasy14lepton}. The lepton cuts increase the value of 
$A_{\phi}$ for the SM from 0.35 to around 0.45, 
and also increase $A_{\phi}$ substantially with anomalous couplings included.
 However, the cuts result in the reduction of signal events and from the Figs. \ref{aziasy7lepton} and \ref{aziasy14lepton}, we 
see that these cuts actually decrease the sensitivity to anomalous couplings.

The azimuthal distribution depends both on 
top polarization and on a kinematic effect. 
According to Eq. \ref{topdecaywidth}, the decay lepton is emitted preferentially along the top 
spin direction in the top rest frame, with $\kappa_f=1$. The
corresponding distributions in the parton cm frame are given by Eq. \ref{angdist} 
with the angular parts described by Eqs. \ref{angmat1} and \ref{angmat2}.
The rest-frame forward (backward) peak corresponds to a peak for
$\cos\theta_{t\ell}=\pm 1$, as seen from the factor $(1 \pm \cos\theta_{t\ell})$ in
the numerator of Eq. \ref{angmat1}. This is the effect of polarization.
The kinematic effect is seen in the factor $(1 - \beta_t \cos
\theta_{t\ell})^3$ in the denominator of Eqs. \ref{angmat1} and
\ref{angmat2}, which again gives rise to peaking for large
$\cos\theta_{t\ell}$. Eq. \ref{costhetatl} therefore implies peaking for
small $\phi_{\ell}$. This is borne out by the numerical results.

\subsection{Energy distribution of charged leptons }
We now study the energy distribution $d\sigma/dE_\ell$ of charged leptons to probe anomalous $tbW$ couplings in 
single-top production and decay. In the rest frame of the top quark, the 
$E_\ell$ distribution of the decay density matrix
$\Gamma(\lambda,\lambda^\prime)$ depends only on the combination of
helicities, $(\lambda,\lambda^\prime)$. To linear order in the couplings,
only the angular part of $\Gamma(\lambda,\lambda^\prime)$ depends on the
helicities, and the energy dependence is the same for all helicity
combinations, and is determined by the effective couplings occurring in
decay. However, there is a weak dependence on the production
differential cross section introduced because 
the boost to the parton cm frame is determined by $\theta_{t\ell}$. 
Thus, the $E_\ell$
distribution arises mainly from the decay process, and depends only
weakly on the polarization.

\begin{figure}[htb]
\begin{center}
\includegraphics[angle=270,width=3.2in]{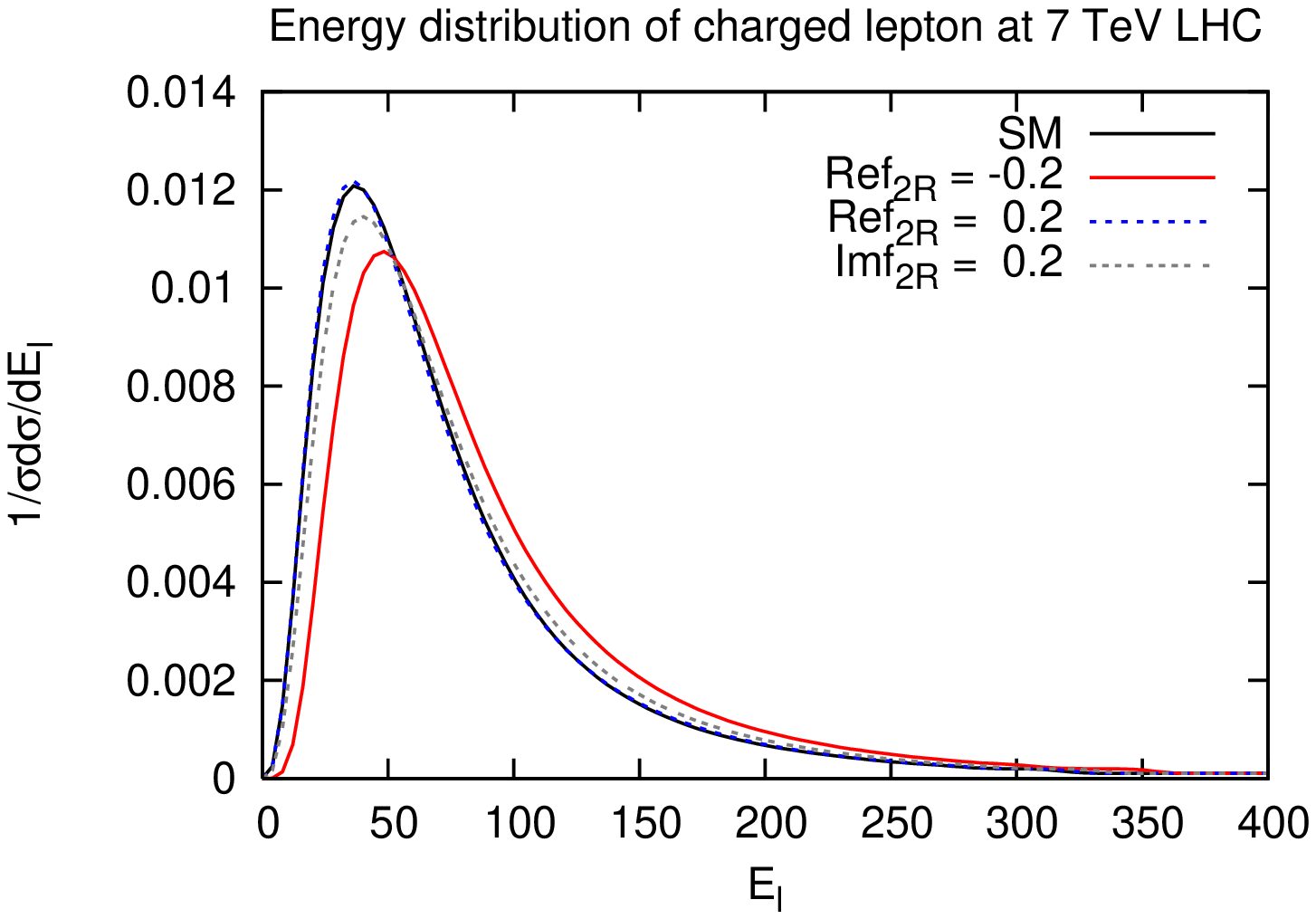} 
\includegraphics[angle=270,width=3.2in]{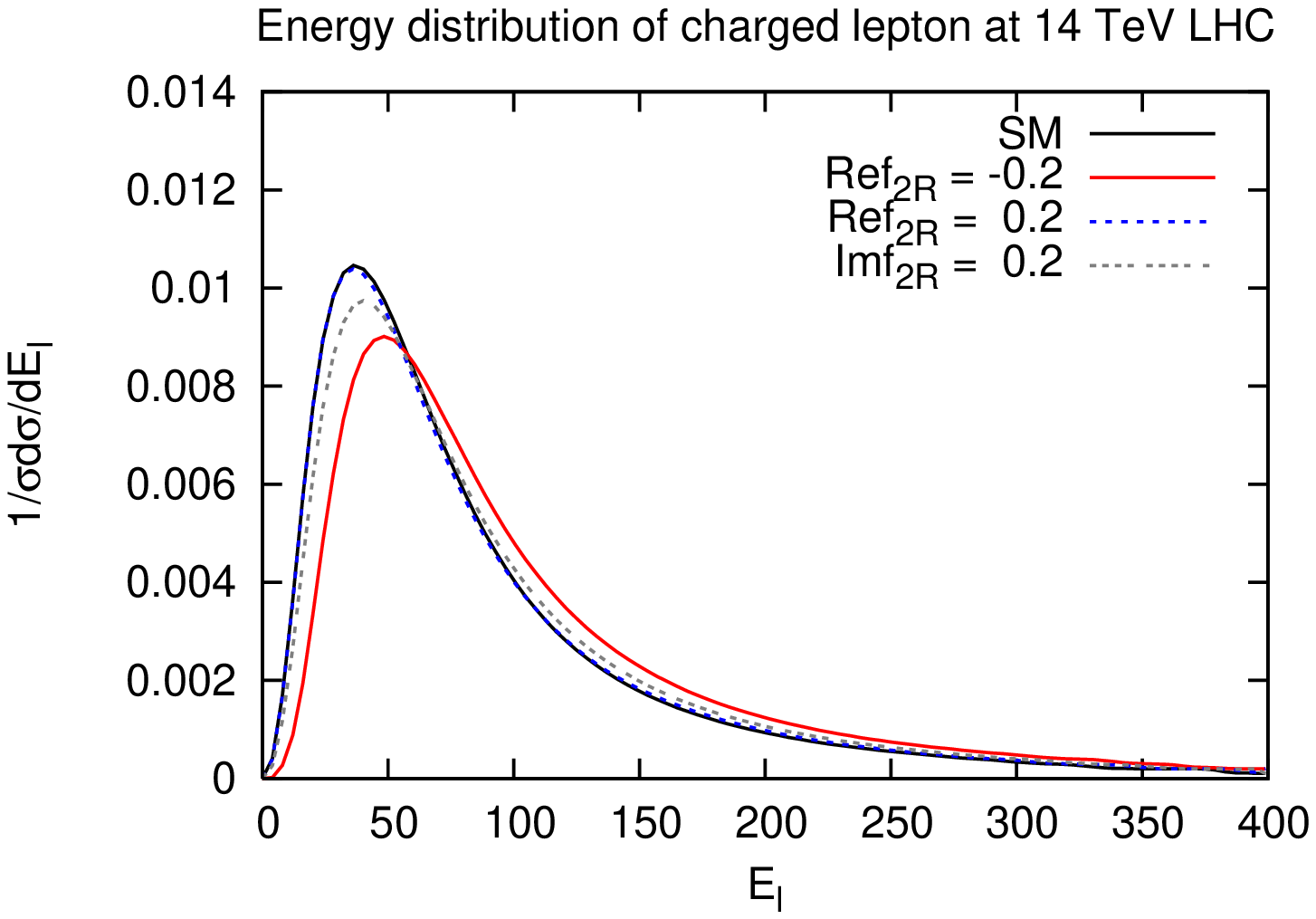}  
\caption{ The energy distribution of the charged lepton in $tW^-$
production at the LHC
for cm energies 7 TeV (left) TeV and 14 TeV (right) for different anomalous $tbW$ couplings.} 
\label{Endistlepton}
\end{center}
\end{figure}

We plot in Fig. \ref{Endistlepton} the $E_\ell$ distribution for 
$\sqrt{s}=$ 7 and 14 TeV.
We see that the distribution is peaked at low values of $E_\ell$ around 40-45 GeV, and all curves 
intersect at a particular value of $E_\ell\approx 62$ GeV. We also observe that the $E_\ell$ distribution is mainly sensitive to $\mathrm{Re}\frb$. 
 
As seen from Fig. \ref{Endistlepton},
the $E_\ell$ distribution is very sensitive to negative values of Re$\frb$ 
and shows little
sensitivity for positive values. The $E_\ell$ distribution at 7 TeV is
peaked slightly more as compared to that for 14 TeV LHC, though the
position of the peak for both  is about the same.

The curves for the $E_\ell$ distribution for anomalous $tbW$ couplings of $\pm 0.2$ and the SM are well separated from each other and 
intersect at $E_\ell^C=62$ GeV. To quantify this difference and to make better  use of statistics, we construct an 
asymmetry around the intersection point of the curves, defined by
\begin{equation}
 A_{E_\ell}=\frac{\sigma(E_\ell
<E_\ell^C)-\sigma(E_\ell>E_\ell^C)}{\sigma(E_\ell <
E_\ell^C)+\sigma(E_\ell > E_\ell^C)},
\label{Enasy}
\end{equation}
\begin{figure}[h]
\begin{center}
\includegraphics[angle=270,width=3.2in]{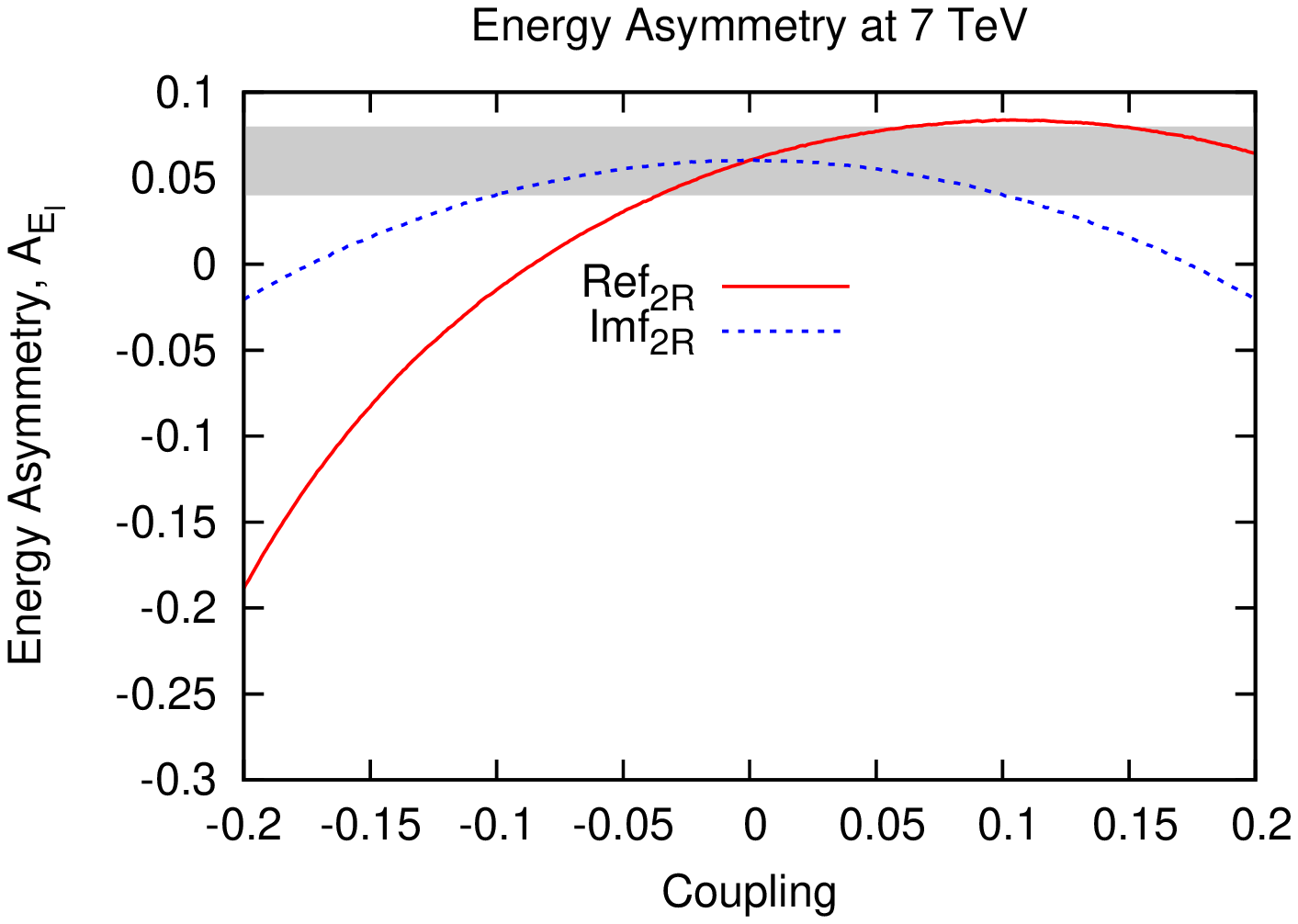} 
\includegraphics[angle=270,width=3.2in]{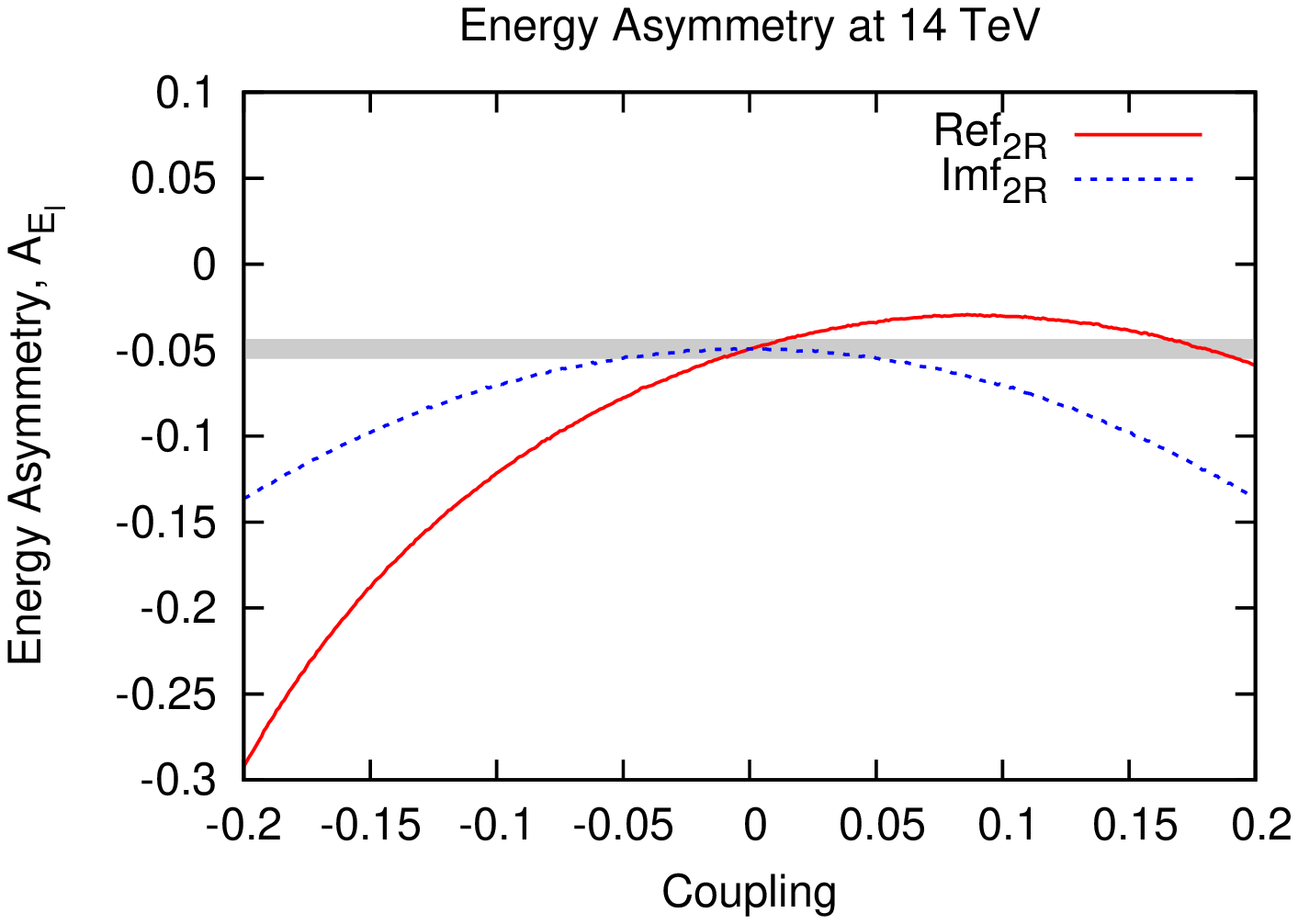} 
\caption{ The energy asymmetry of the charged lepton in $tW^-$
production at the LHC
for cm energy 14 TeV without lepton cuts (left) and with cuts (right), as a function of anomalous $tbW$ couplings.
The grey band corresponds to the energy asymmetry
predicted in the SM with a 1$\sigma$ error interval. } 
\label{Enasylepton}
\end{center}
\end{figure}
where the denominator is the total cross section. Plots for $A_{E_\ell}$
as a function the coupling
are shown in Fig. \ref{Enasylepton} for two cm energies of 7 TeV and 14
TeV.  We can see from the figure that $A_{E_\ell}$ is 
very sensitive to $\mathrm{Re}\frb$ and hence can be a sensitive probe
of this coupling. It is also seen from the figure that the $A_{E_\ell}$
for the SM is positive for $\sqrt{s}=7$ TeV, but negative for
$\sqrt{s}=14$ TeV. This is in accordance with a sharper peaking of the
energy distribution in the former case. Another difference between the
asymmetries for the two cm energies is that $A_{E_\ell}$ changes sign
with the sign for some value of $\mathrm{Re}\frb$ in case of
$\sqrt{s}=7$ TeV, but remains negative in case of $\sqrt{s}=14$ TeV. 

\section{\boldmath Angular distribution of $b$ quarks }
Although the charged-lepton azimuthal distribution provides a neat way to probe top polarization 
independent of new physics in the top-decay vertex, it suffers from low branching ratio of $W$ and hence 
low number of events for the analysis.  This situation can be improved
upon by using $b$-quark angular distributions, without restricting only
to the leptonic decays of the $W$ coming from top decay,
and thus utilizing all the single-top events. We thus assume, for
purposes of this section, that the top quark can be identified in 
hadronic and semi-leptonic decays with sufficiently 
good efficiency to enable measurement of
$b$-quark distribution in all of them.

As described earlier, we use NWA to factorize the full process into single-top production and 
top decay. Similar to Eq. \ref{dsigell}, we can write the full differential cross section for the process 
$g(p_g)b(p_b)\rightarrow t(p_t,\lambda_t)W^-$ followed by $t(p_t,\lambda_t)\rightarrow b(p_b')W^+$ as 
\bea\label{bdist}
\frac{d\sigma}{d\cos_{\theta_t}d\Omega_b}&=&
\frac{1}{128(2\pi)^3\shat^{3/2}}
\frac{|\overrightarrow{p_t}|}{\Gamma_t
m_t} 
\frac{(m_t^2 - m_W^2)}{E_t^2(1-\beta_t\cos\theta_{bt})^2}
\sum_{\lambda_t,\lambda_t^\prime}\left[\rho(\lambda_t,\lambda_t^\prime)\, \Gamma(\lambda_t,\lambda_t^\prime)\right],
\eea
where the polar angle $\theta_t$ of the top quark, and the polar and
azimuthal angles $\theta_b$ and $\phi_b$  of the $b$ quark produced in
top decay, are measured with respect to the parton direction as the $z$
axis, and with the $xz$ plane defined as the plane containing the top
momentum. $\theta_{bt}$ is the angle between the top momentum and the
momentum of the decay $b$ quark.  

The density matrix for single-top production $\rho(\lambda_t,\lambda_t^\prime)$ appearing in Eq. \ref{bdist} is given in the Appendix. 
The decay density matrix for $t\rightarrow bW$ in the top-quark rest frame is given by \footnote{The expressions for charged-lepton and 
$b$-quark angular distributions agree with those given in Ref. \cite{AguilarSaavedra:2010nx}}
\bea
\Gamma(\pm,\pm)&=&\frac{g^2 m_t^2}{2}\left[\mathcal C_1\pm\mathcal C_2
\cos\theta_b^0\right],\\
\Gamma(\pm,\mp)&=&\frac{g^2 m_t^2}{2}\left[\mathcal C_2 \sin\theta_b^0
e^{\pm\phi_b^0}\right],
\eea
where
\bea
\mathcal C_1&=&\frac{1}{2r^2}(1-r^2)\left[|\fla|^2(2r^2+1)
+ \mathrm{Re}\fla\mathrm{f^*_{2R}}\,6r 
+ \lvert\frb\rvert^2 (2+r^2) \right],\\
\mathcal C_2&=&\frac{1}{2r^2}(1-r^2)\left[|\fla|^2(2r^2-1)
+ \mathrm{Re}\fla\mathrm{f^*_{2R}}\,2r 
+ \lvert\frb\rvert^2 (2-r^2) \right].
\eea
The rest frame polar and azimuthal angles of the $b$, respectively
$\theta_b^0$ and $\phi_b^0$, may be expressed in terms of the parton cm
frame angles in a straightforward way.

\begin{figure}[htb]
\begin{center}
\includegraphics[angle=270,width=3.2in]{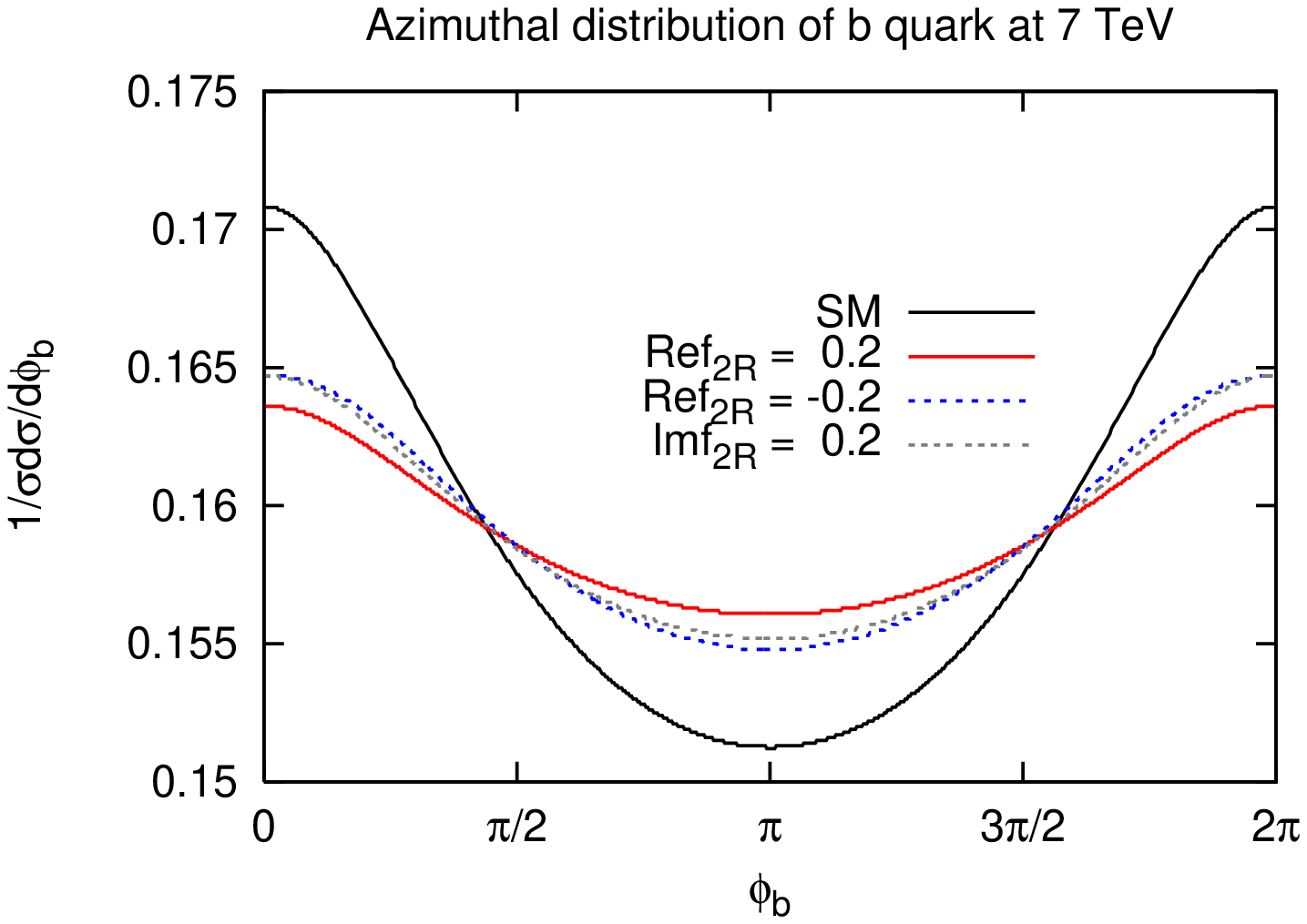} 
\includegraphics[angle=270,width=3.2in]{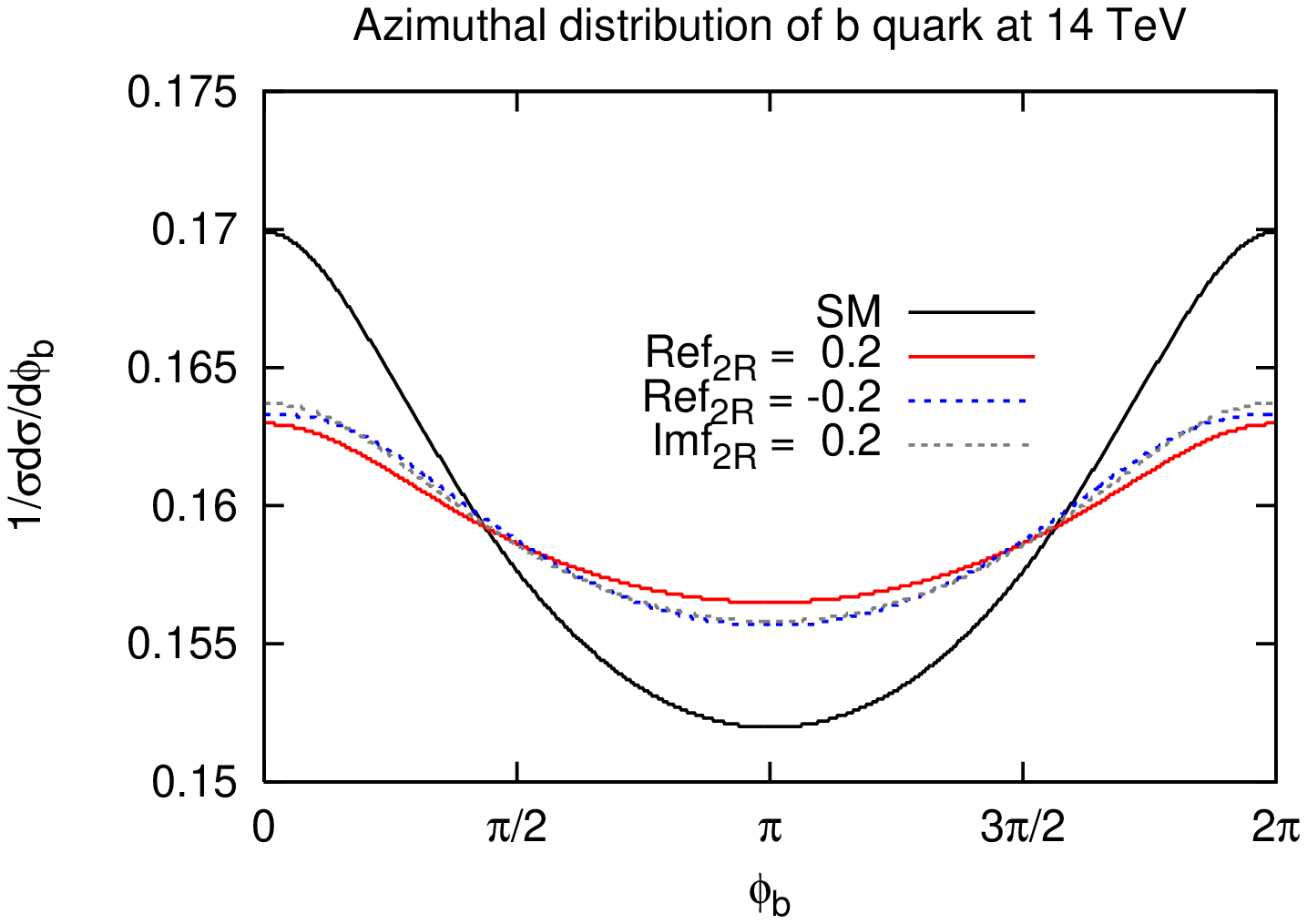} 
\caption{ The azimuthal distribution of the $b$ quark in $tW^-$ production
at the LHC for cm energies 7 TeV and 14 TeV for different anomalous $tbW$ couplings.} 
\label{distazibquark}
\end{center}
\end{figure}

Plots for the azimuthal distribution of the $b$ quark for different values of anomalous couplings $\mathrm{Re}\frb$ 
and $\mathrm{Im}\frb$ are shown in Fig. \ref{distazibquark}. The curves for  values $\pm 0.2$ of these couplings are well separated from 
each other and from the SM curve. In the azimuthal distribution of the $b$ quark, we get dependence  on 
anomalous couplings both from production as well as decay. We find that the contributions from production 
and from decay come with opposite signs, partially cancelling each other.

To study the sensitivity and make the best use of azimuthal $b$-quark
distribution, we construct an asymmetry $A_{b}$ : 
\begin{equation}
 A_{b}=\frac{\sigma(\cos \phi_b >0)-\sigma(\cos \phi_b<0)}{\sigma(\cos
\phi_b >0)+\sigma(\cos \phi_b<0)}.
\label{aziasyb}
\end{equation}
\begin{figure}[h]
\begin{center}
\includegraphics[angle=270,width=3.2in]{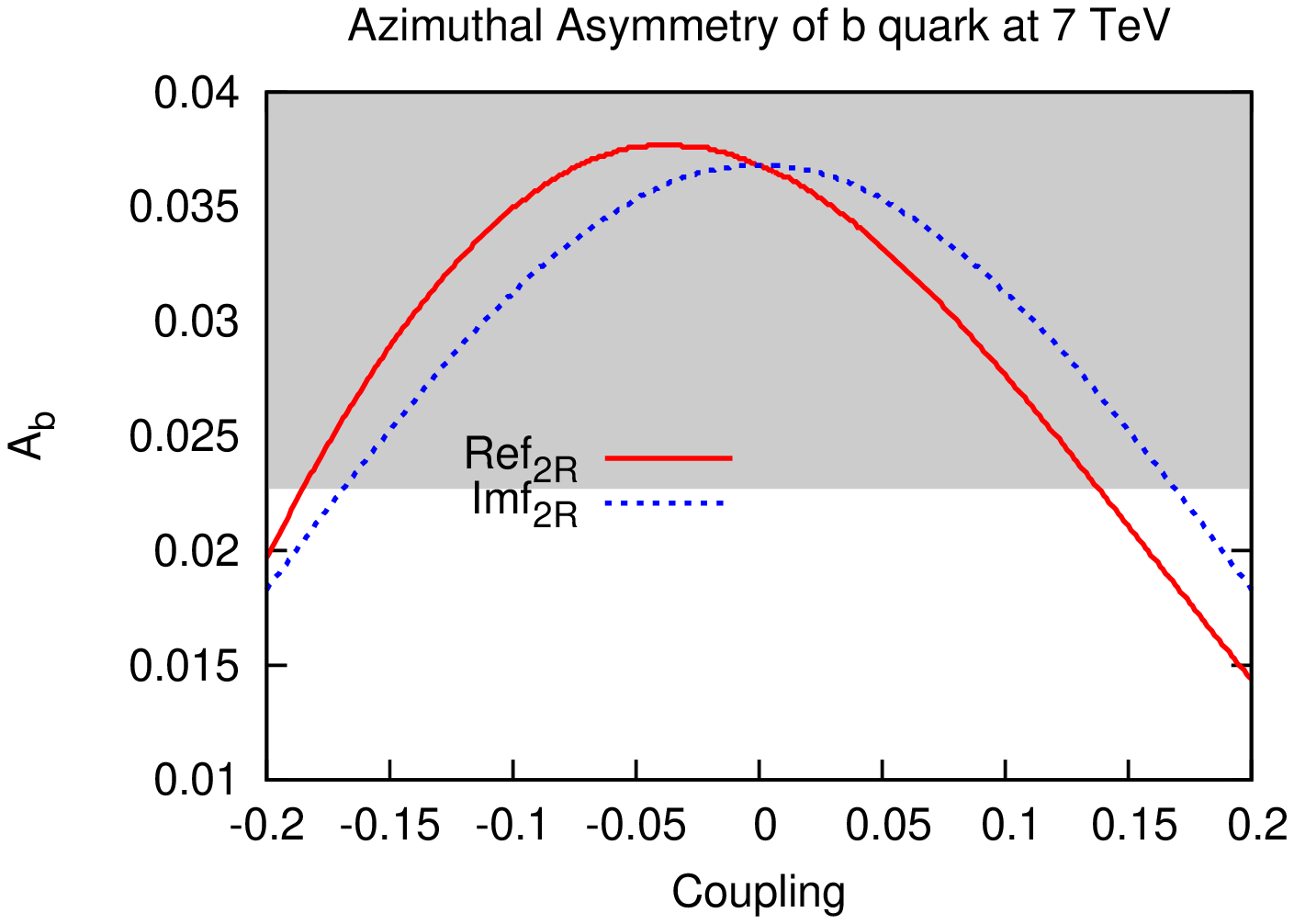} 
\includegraphics[angle=270,width=3.2in]{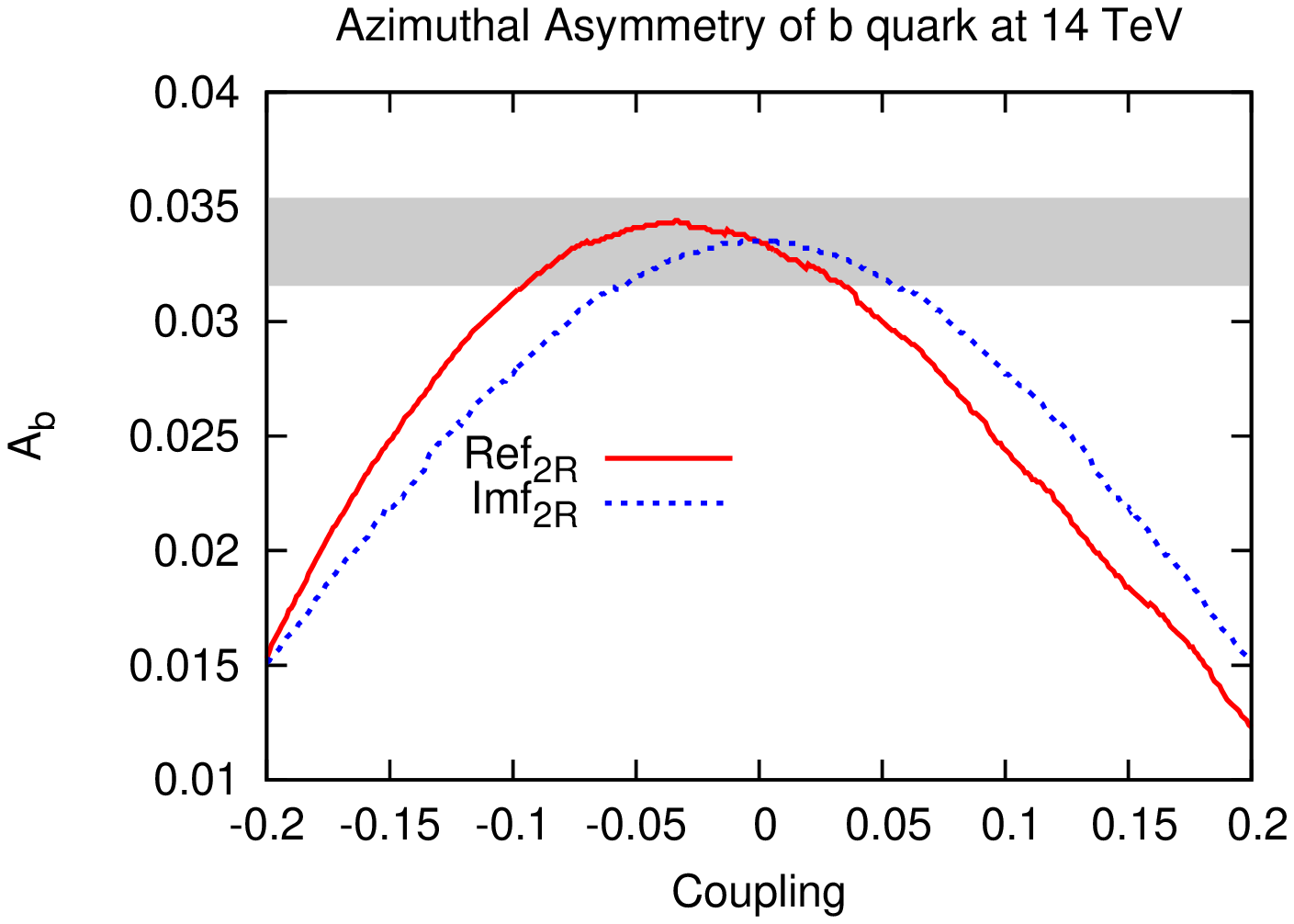} 
\caption{ The azimuthal asymmetry of the $b$ quark in
$tW^-$ production at the LHC
for cm energies 7 TeV and 14 TeV as a function of anomalous $tbW$ couplings.
 The grey band corresponds to the asymmetry predicted in the SM with
a 1$\sigma$ error interval.} 
\label{aziasymbquark}
\end{center}
\end{figure}

 The asymmetry $A_b$ is plotted in Fig. \ref{aziasymbquark} as a function of anomalous couplings
for the cm energies $\sqrt{s}=$ 7 TeV and 14 TeV. From the figure it is clear that $A_b$ shows less sensitivity to 
couplings $\mathrm{Re}\frb$ and Im$\frb$ as compared to other asymmetries. As stated earlier, this is due to the fact that 
the contributions of anomalous couplings to the asymmetry from the production and the decay are of opposite in sign 
and hence tend to cancel each other.

\section{\boldmath Sensitivity analysis for anomalous $tbW$
couplings}\label{sens}
We now study the sensitivities of the observables discussed in the 
previous sections to the anomalous $tbW$ couplings at the LHC running at
two cm of energies viz., 7 TeV and 14 TeV, 
with integrated luminosities 1 fb$^{-1}$ and 10 fb$^{-1}$, respectively. To obtain the 1$\sigma$ limit on the anomalous 
$tbW$ couplings from a measurement of an observable, we find those
values of the 
couplings for which observable deviates by 1$\sigma$ from its SM value. The statistical uncertainty $\sigma_i$
in the measurement of any generic asymmetry $\mathcal A_i$ is given by 
\beq\label{stat-dev}
\sigma_i=\sqrt{\frac{1-\left({\mathcal A}^{SM}_i\right)^2}{\mathcal N}},
\eeq
where ${\mathcal A}^{SM}_i$ is the asymmetry predicted in the SM and $\mathcal N$ is the total number of events  
predicted in the SM. 
We apply this to the various asymmetries we have discussed. In case of the top polarization asymmetry, the limits are obtained on the
assumption that the polarization can be measured with 100\% accuracy.

The 1$\sigma$ limits on Re$\frb$ and Im$\frb$ are given in Table \ref{lim14TeV} where we assume only 
one anomalous coupling to be non-zero at a time. We have also assumed measurements on a $tW^-$ final state. Including the $\bar t W^+$ final
state will improve the limits by a factor of $\sqrt{2}$. In case of the lepton distributions, we take into account only one leptonic channel.
Again, including other leptonic decays of the top would improve the limits further. The limits
corresponding to a linear approximation in the couplings are denoted by
the label ``lin. approx.". Apart from the 1$\sigma$ limits shown in
Table \ref{lim14TeV}, which correspond to intervals which include zero
value of the coupling, there are 
other disjoint intervals which could be ruled out by null if no
deviation from the SM is observed for $P_t$ and
$A_{E_\ell}$. This is apparent from Figs. 
\ref{polcoup} and \ref{Enasylepton}. The additional allowed intervals
for Re$\frb$ from $P_t$ measurement are [0.158, 0.205] and
[0.160, 0.167] for cm energies of 7 TeV and 14 TeV, respectively.
The corresponding intervals for $A_{E_\ell}$ are [0.147, 0.285] and [0.175, 0.185] 
\footnote{[a, b] denotes the allowed values of the coupling $f$ at the
1$\sigma$ level, satisfying $a<f<b$.}.

 \begin{table}[h]
  \begin{tabular}{c|c|c|c|c}
   \hline
&\multicolumn{2}{c|}{7 TeV}&\multicolumn{2}{c}{14 TeV}\\
\hline
\hline
 Observable & 	$\mathrm{Re}\frb$	&  $\mathrm{Im}\frb$ & 	$\mathrm{Re}\frb$	&  $\mathrm{Im}\frb$ \\
 \hline
 $P_t$ 				&[$-0.025$, $0.032$]	& [$-0.072$, $0.072$] 		&[$-0.004$, $0.004$]	& [$-0.034$, $0.034$] \\
 $P_t$ (lin. approx.) 		&[$-0.027$, $0.027$]	& $-$				&$[-0.004$, $0.004]$	&  $-$\\
 $A_\phi$ 			&[$-0.133$, $0.194$]	& [$-0.150$,$0.150$] 		&[$-0.034$, $0.086$]	& [$-0.050$, $0.050$] \\
 $A_{\phi}$ (lin. approx.)	&[$-0.204$, $0.204$]	& 	$-$			&[$-0.030$, $0.030$]	&  $-$\\
 $A_b$ 				&[$-0.191$, $0.147$]	& [$-0.177$, $0.177$] 		&[$-0.096$, $0.035$]	& [$-0.059$, $0.059$] \\
 $A_{E_{\ell}}$ 		&[$-0.044$, $0.073$]	& [$-0.114$, $0.114$]		&[$-0.006$, $0.009$]	& [$-0.038$, $0.038$]\\
\hline
\hline
  \end{tabular}
\caption{Individual limits on real and imaginary parts of anomalous coupling $\frb$ which may be obtained by the measurement of the observables 
shown in the first column of the table at two cm of energies viz., 7 TeV and 14 TeV with integrated luminosities of 1 fb$^{-1}$ and 10
fb$^{-1}$ respectively. A dash ``$-$'' indicates that no limits are possible. }
 \label{lim14TeV}
 \end{table}

It is seen that the azimuthal asymmetry $A_{\phi}$ and the energy
asymmetry $A_{E_{\ell}}$ of the charged lepton are more 
sensitive to 
negative values of the anomalous couplings $\mathrm{Re}\frb$.
$A_{E_{\ell}}$ is the most sensitive of the asymmetries we
consider. 
In fact, the sensitivity of $A_{E_{\ell}}$ to $\mathrm{Re}\frb$ and Im$\frb$ is comparable 
to the sensitivity of top polarization to the same couplings, despite the
fact that only one leptonic decay channel, with a branching fraction of
about 1/9, is considered for
 $A_{E_{\ell}}$. 
The additional contribution to $A_{E_{\ell}}$ of the $\frb$ coupling through the top
decay channel seems to compensate for the low branching fraction. 
$A_{b}$ is seen to have the lowest sensitivity, where there is partial
cancellation of contributions to the asymmetry from production and from
decay.

We also obtain simultaneous limits (taking both $\mathrm{Re}\frb$ and $\mathrm{Im}\frb$ non-zero simultaneously) 
on these anomalous couplings that may be obtained by combining the measurements of all observables. 
\begin{figure}[h]
\begin{center}
\includegraphics[width=2.0in]{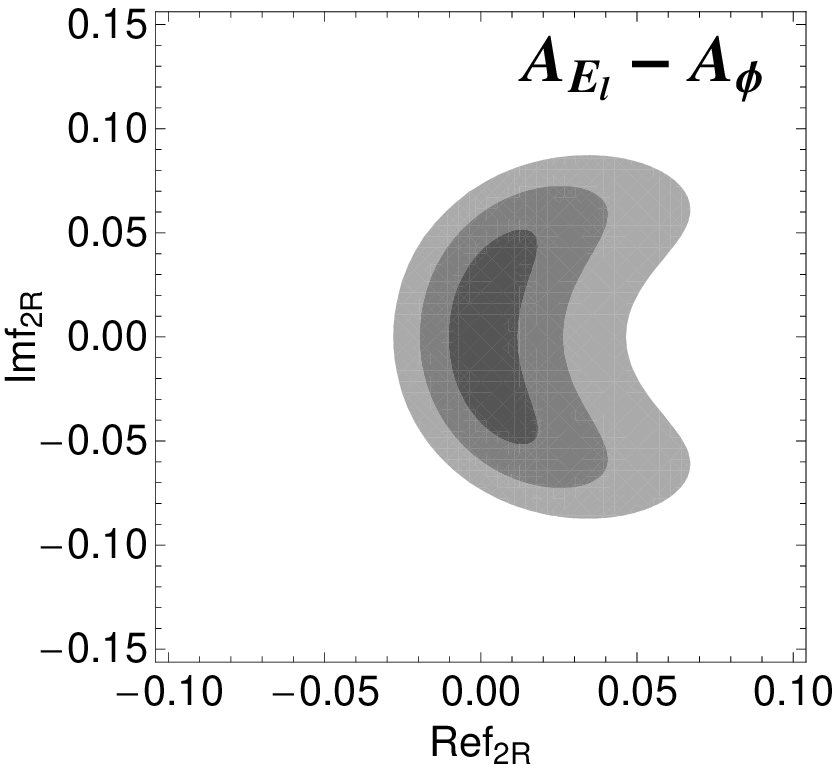} 
\includegraphics[width=2.0in]{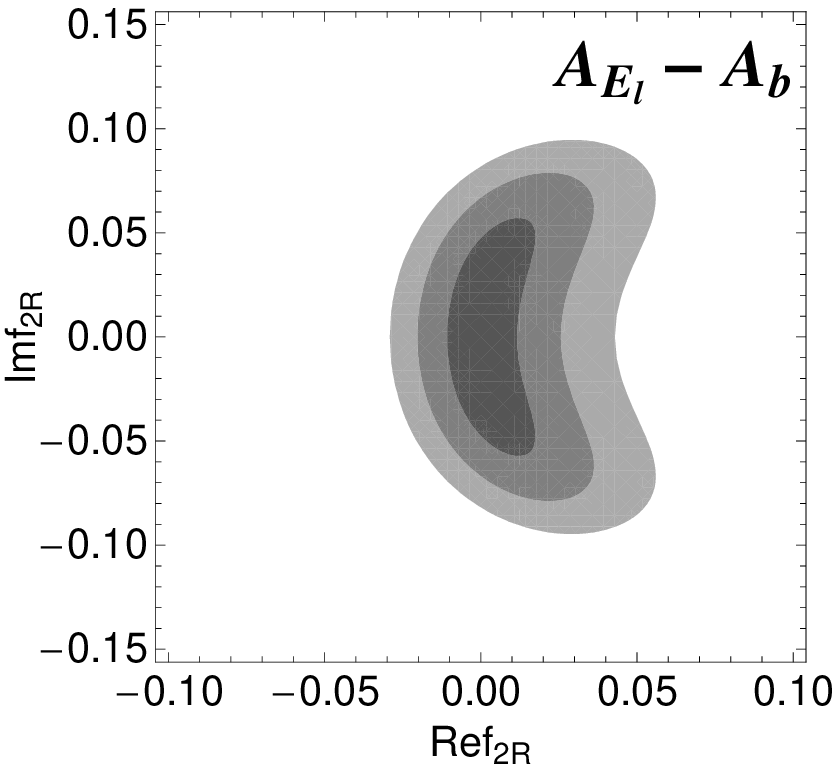} 
\includegraphics[width=2.0in]{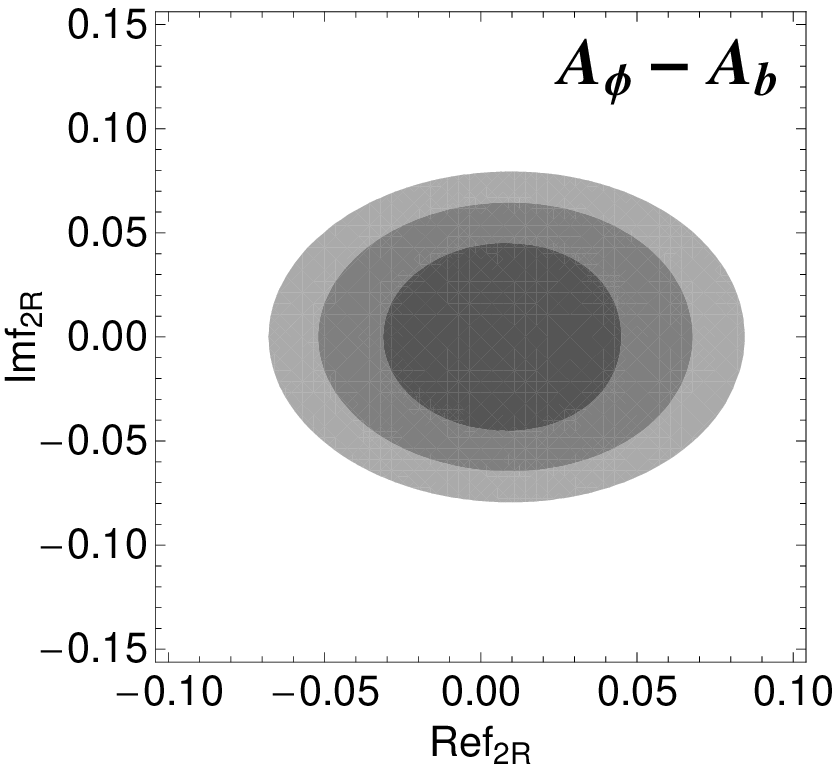} 
\caption{The 1$\sigma$ (central region), 2$\sigma$ (middle region) and 3$\sigma$ (outer region) CL regions in the 
$\mathrm{Re}\frb$-$\mathrm{Im}\frb$ plane allowed by the combined measurement of two observables at a time. The left, centre and right plots 
correspond to measurements of the combinations $A_{E_\ell}$-$A_\phi$,
$A_{E_\ell}$-$A_b$ and $A_\phi$-$A_b$ respectively.  The $\chi^2$ values for 1$\sigma$, 2$\sigma$ and 3$\sigma$ CL intervals 
are 2.30, 6.18 and 11.83 respectively for 2 parameters in the fit.}
\label{lim-ind}
\end{center}
\end{figure}
For this, we perform a $\chi^2$ analysis to fit all the observables to within $f\sigma$ of statistical errors 
in the measurement of the observable. We define the following $\chi^2$ function 
\beq\label{chisq}
\chi^2 = \sum_{i=1}^n\left(\frac{P_i-O_i}{\sigma_i}\right)^2,
\eeq
where the sum runs over the $n$ observables measured and $f$ is the
degree of the confidence interval. $P_i$'s are the values of the observables
obtained by taking both anomalous couplings non-zero (and is a function of 
the couplings $\mathrm{Re}\frb$ and $\mathrm{Im}\frb$) and $O_i$'s are the values of the observables 
obtained in the SM. $\sigma_i$'s are the statistical fluctuations in the measurement of the observables, 
given in Eq. \ref{stat-dev}.

 In Fig. \ref{lim-ind}, we show the 1$\sigma$, 2$\sigma$ and 3$\sigma$ regions in $\mathrm{Re}\frb$-$\mathrm{Im}\frb$ 
plane allowed by combined measurement of asymmetries $A_{E_\ell}$, $A_\phi$ and $A_b$ taken two at a time. 
For this, in the $\chi^2$ function of Eq. \ref{chisq}, we have taken only two of the three observables at a time. 
From among the three combinations shown in Fig. \ref{lim-ind}, we find that the strongest simultaneous limits come from the 
combined measurement of $A_{E_\ell}$ and $A_\phi$, viz.,  [-0.01, 0.02]  on
$\mathrm{Re}\frb$  and [$-0.05$, 0.05] on $\mathrm{Im}\frb$,  at the $1\sigma$ level.
\begin{figure}[h]
\begin{center}
\includegraphics[width=3.0in]{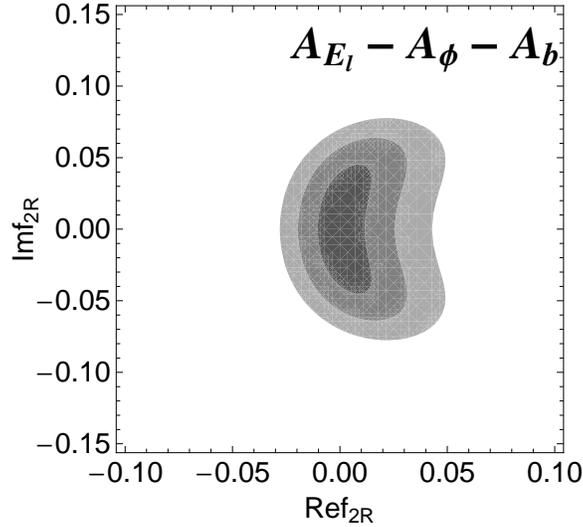} 
\caption{The 1$\sigma$ (central region), 2$\sigma$ (middle region) and 3$\sigma$ (outer region) CL regions in the 
$\mathrm{Re}\frb$-$\mathrm{Im}\frb$ plane allowed by the combined measurement of all the observables simultaneously.} 
\label{lim-all}
\end{center}
\end{figure}

In Fig. \ref{lim-all}, we show the 1$\sigma$, 2$\sigma$ and 3$\sigma$ regions in $\mathrm{Re}\frb$-$\mathrm{Im}\frb$ 
plane allowed by combined measurement of all three asymmetries $A_{E_\ell}$, $A_\phi$ and $A_b$ 
simultaneously. We find that the combined measurement of all the observables provide the most stringent simultaneous limits on 
$\mathrm{Re}\frb$ and $\mathrm{Im}\frb$ of [-0.010, 0.015] and $[-0.04$, 0.04] respectively at 1$\sigma$. 
We find that the energy asymmetry $A_{E_\ell}$ plays a crucial role in determining the combined limits.

 We now compare our results with those of other works 
on the determination of anomalous $tbW$ couplings at the LHC. 
Refs. \cite{Boos:1999dd,AguilarSaavedra:2008gt,AguilarSaavedra:2010nx,Najafabadi:2008pb,Espriu:2001vj,Espriu:2002wx,AguilarSaavedra:2011ct}, 
have studied single-top production at the LHC in the context of anomalous $tbW$ couplings. Boos et al. \cite{Boos:1999dd} 
find a limit of $−0.12 < \frb < 0.13$ at the LHC with 100 fb$^{-1}$ of luminosity. Refs. \cite{Espriu:2001vj,Espriu:2002wx}  
considered couplings $\fla$ and $\fra$ in their analysis and ignored $\frb$ on which we focus. In Ref. \cite{AguilarSaavedra:2008gt}, the authors 
have studied all three single-top production channels to probe anomalous $tbW$ couplings and have utilized combinations of observables like 
cross sections, $W$ polarization helicity fractions in top decay, and other angular asymmetries. They predict a 1$\sigma$ limit of [$-0.012$, 0.024] on the 
coupling $\frb$ with an integrated luminosity of 30 fb$^{-1}$. Ref. \cite{AguilarSaavedra:2010nx} determine expected 3$\sigma$ limits 
on Re$\frb$ to be $[-0.056$, 0.056] and on Im$\frb$ to be $[-0.115$,
0.115], with 10 fb$^{-1}$ of integrated luminosity. 
Najafabadi \cite{Najafabadi:2008pb} has studied 
the $tW$ channel for single-top production and determined the expected 1$\sigma$ CL limits on anomalous coupling $\frb$ to be in the range 
[$-0.026$, 0.017] with 20 fb$^{-1}$ of integrated luminosity at 14 TeV LHC using the single-top production cross section. 

Refs. \cite{Hubaut:2005er,AguilarSaavedra:2007rs,AguilarSaavedra:2006fy} have studied various observables in 
$t\bar{t}$ pair production at the LHC with semileptonic decays of the top. 
Ref. \cite{AguilarSaavedra:2006fy} predict 1$\sigma$ limit on $\frb$ of $[-0.019, 0.018]$ with 10 fb$^{-1}$ of integrated luminosity. 
Refs. \cite{AguilarSaavedra:2007rs, Hubaut:2005er} study $W$ polarization in 
top decay in top-quark pair production at the LHC to constrain the anomalous $tbW$ couplings. 
They construct various asymmetries and helicity fractions to probe anomalous couplings in the decay of the top quark.  Ref. \cite{ Hubaut:2005er} 
quotes a 2$\sigma$ limit of $0.04$ on $\frb$ with full detector-level simulations including systematic uncertainties and with 
different observables. Ref. \cite{AguilarSaavedra:2007rs} obtain a limit
on $\frb$ of [$-0.0260$, 0.0312] with simulation of the ATLAS detector. 

All these analyses except those of \cite{AguilarSaavedra:2006fy,AguilarSaavedra:2010nx} consider anomalous $tbW$ couplings to be real parameters. 
In our analysis, we consider all anomalous couplings to be complex
and find that only the real part of the 
coupling $\frb$ gives significant contribution to all observables at linear order in anomalous couplings. 
Without a linear approximation, other couplings also contributes at the
quadratic level. However, we focus only on the $\frb$ since its contribution is
dominant, occurring as it does at linear order. With integrated
luminosity 10 fb$^{-1}$ and cm energy 14 TeV, we find the most stringent 
limits possible on $\mathrm{Re}\frb$ to be $[-0.006,0.009]$ and on $\mathrm{Im}\frb$ to be $\pm 3.8\times 10^{-2}$, coming from the lepton
energy asymmetry $A_{E_\ell}$, which are nominally an order of magnitude better than the Tevatron direct search limit and 
 better than the limits obtained in Refs. \cite{Boos:1999dd, AguilarSaavedra:2007rs, AguilarSaavedra:2008gt, 
AguilarSaavedra:2006fy, Hubaut:2005er, AguilarSaavedra:2010nx}. Our estimate [$-0.044$, 0.073] for limits on Re$\rm f_{2R}$ in the $\sqrt{s}=7$ TeV 
run is comparable to the numbers obtained by the extrapolation of the result of \cite{AguilarSaavedra:2011ct} to an integrated luminosity of 1 fb$^{-1}$.
It is of course true that including realistic detector efficiencies,
especially for $b$ tagging, will worsen our limits somewhat. But, the crucial point in our analysis is that we are 
able to determine limits on real and imaginary parts of coupling $\frb$ separately while others determine limits only on the 
magnitude of $\frb$.

\section{Backgrounds and next-to-leading-order corrections}

It is worthwhile to examine the dominant backgrounds to our signal process $gb\rightarrow tW^-$. Background estimation 
and extraction of the signal for this process has been studied in detail in Refs. \cite{Tait:1999cf,ATL-rep}. 
The main background for this signal would come from (a) processes which contain continuum of $W^+W^-b$ 
involving an off-shell top quark, (b) top-pair production where one of
the $b$ quarks is missed as 
it lies outside the detector range, (c) processes containing $W^+W^-j$ 
where lighter-quark jet $j$ is misidentified as a $b$-quark jet (the
probability being 1\%). The contributions of the processes $W^+W^-b$ and
$W^+W^-j$,
which are of order $O(\alpha_s \alpha_W^2)$, 
are much 
smaller than the $tW^-$ signal, which is 
of order $O(\alpha_s\alpha_W)$. Tait \cite{Tait:1999cf} has considered both W's to decay leptonically 
and hence the final state consists of two hard charged leptons $+$ one
$b$ jet + missing $E_T$ (arising from two neutrinos). 
For such a signal, the process $ZZj$ would also act as background where one of the $Z$ decays into a pair of 
the charged leptons and the other $Z$ decays into neutrinos. In Ref. \cite{Tait:1999cf}, all the backgrounds are 
simulated at LO in the strong and weak couplings and standard acceptance cuts ($p_T>15$ GeV and $\eta < 2$) 
are applied on all final state particles. 
To suppress large $t\bar{t}$ background, it is required that there should not be
more than one hard $b$ jet. After applying these cuts and with integrated luminosity less than 1 fb$^{-1}$ at 14 TeV LHC, the conclusion 
of the Ref. \cite{Tait:1999cf} is that 5$\sigma$ observation of single-top events in $tW^-$ channel is possible.

The authors of Ref. \cite{ATL-rep} consider the situation where the $W$ coming
from the top quark decays leptonically and the other $W$ 
decays into two light-quark jets. Therefore the signal would consist of
three jets, one of which is a hard $b$ jet, 
an isolated hard charged lepton and missing energy. 
The jet multiplicity requirement rejects a major part of the 
$t\bar{t}$ background. 
Also, the requirement of the two-jet invariant mass to be within the vicinity of
the $W$ mass (70 GeV-90 GeV) eliminates all backgrounds 
which do not have another $W$, as for example $W+$jets, other single-top
and QCD processes. In all these analyses, $b$-tagging efficiency 
is assumed to be 60\%. After applying all the cuts, it has been shown that 10\% sensitivity can be achieved with 1 fb$^{-1}$ by 
combining both electron and muon channels.

Turning to radiative corrections,
NLO QCD corrections to the process $gb\rightarrow tW^-$ in the context
of the 14 TeV LHC has been studied in detail in 
Ref. \cite{Zhu:2001hw}. These corrections are substantial, up to 70\% of
the LO cross section. They are shown to be 
dependent on the factorization scale and increase steadily with the increment in the scale. For factorization scale 
$\mu=\mu_0/2=(m_t+m_W)/2$, the $K$ factor for the QCD correction is 1.4 while for $\mu=2\mu_0=2(m_t+m_W)$, it is around 1.7. In 
our analysis, we have taken $\mu=m_t$, for which 
the $K$ factor is expected to be about 1.5. 

The complete NLO EW corrections have been calculated in Ref. \cite{Beccaria:2007tc} for $pp\rightarrow tW^-+X$ in the 
context of the LHC. The EW corrections are always positive and are maximum for $tW$ invariant mass closed to threshold value. 
With the increase in $tW$ invariant mass, these EW corrections decrease. So, the maximum EW correction is around 7\%  at 
threshold and it goes down to 3.5\% for $tW$ invariant mass of 1200 GeV.

In our analysis, we have not included any $K$ factor.
Including NLO factors in our analysis would not change the asymmetries
much, but
would increase the signal and hence, the sensitivity 
on anomalous couplings would be enhanced.

\section{Summary and conclusions}

We have investigated the sensitivity of the LHC to anomalous $tbW$ couplings in single-top production in association with a 
$W^-$ boson followed by semileptonic decay of the top. We derived analytical expressions
for the spin density matrix of the top quark including  
contributions of both real and imaginary parts of the anomalous $tbW$ couplings. We find that in the limit of vanishing 
$b$ quark mass, only the real part of coupling $\frb$ contributes to the spin density matrix at linear order. Because of 
the chiral structure of the anomalous $tbW$ couplings, the resulting top polarizations are vastly different from those expected 
in the SM. We find that substantial deviations, as much as 20-30\%, in the degree of longitudinal top polarization from 
the SM value of $-0.256$ are possible even for anomalous couplings of
magnitude 0.1. The degree of longitudinal top polarization varies from
$-0.075$ to $-0.363$ for $\mathrm{Re}\frb$ while for Im$\rm f_{2R}$ it is symmetric around the SM value and varies from $-0.139$ to $-0.139$ 
with coupling varying from $-0.2$ to 0.2 as compared to the SM value of $-0.256$ for 14 TeV LHC. 

Since top polarization can only be measured through the distributions of its decay products in top decay, 
we studied distributions of top decay products. We consider the top to decay semi-leptonically, since this channel is
expected to have the best accuracy and spin analyzing power. However, decay distributions can get contributions from anomalous
couplings responsible for top polarization as well as for top decay.
We find that normalized charged-lepton azimuthal and energy distributions and $b$-quark azimuthal distributions
are sensitive to anomalous couplings $\mathrm{Re}\frb$ and $\mathrm{Im}\frb$. In each case, we define an asymmetry, whose deviation
from the SM value would be a measure of the anomalous couplings. We find that the azimuthal asymmetry $A_{\phi}$ and the energy
asymmetry $A_{E_{\ell}}$ of the charged lepton are more sensitive to 
negative values of the anomalous couplings $\mathrm{Re}\frb$. A limit of [$-0.034$, 0.086] on $\mathrm{Re}\frb$ would be 
possible from $A_{\phi}$ for cm energy of 14 TeV at the LHC with integrated luminosities of 10 fb$^{-1}$ respectively. 
For $\mathrm{Im}\frb$ the corresponding limit is $\pm 0.050$. 
$A_{E_{\ell}}$, which is the most sensitive of the asymmetries we consider, probes Re$\frb$ and Im$\frb$ in the ranges
[$-0.006$, 0.009] and $\pm 0.038$ respectively. Limits from $A_{b}$ are the least stringent, though not much worse than
those from $A_{E_{\ell}}$.

The above results correspond to assuming only one coupling to be nonzero at a time.
We also estimated simultaneous limits on $\mathrm{Re}\frb$ and $\mathrm{Im}\frb$ by 
combining measurements of all observables using a $\chi^2$ analysis. The best possible 1$\sigma$ limits on these couplings were found to be  
[$-0.010$, 0.015] and [$-0.050$, 0.050], respectively.

The limits we estimate for LHC with $\sqrt{s}=7$ TeV and integrated luminosity 1 fb$^{-1}$ are obviously worse. Nevertheless, 
they are comparable to those expected from an analysis of $W$ helicities as carried out in \cite{AguilarSaavedra:2011ct}.

In summary, our proposal will enable limits to be placed on $\frb$ which are somewhat better than 
limits expected from other measurements at the LHC, and at least an order of magnitude better than the indirect limits.

Our results would be somewhat worsened by the inclusion realistic detection efficiencies for the $b$ jet and for the detection of the
$W$. However, we would like to emphasize that since we do not require accurate reconstruction of the full top-quark four-momentum, the limits are not
likely to be much worse. On the other hand, inclusion of the $\bar tW^+$ final state, as well as additional leptonic channels in top decay would
contribute to improving on our estimates of the limits. A more complete analysis including detector simulation would be worthwhile to carry
out.
\section*{Appendix}

In this appendix, we give the spin density matrix elements for the single-top production process. 
For this we use the following notation for scalar products of four-momenta involved in the process:
$(\ptpb)=\mathcal S_{tb}$, $(\ptpg)=\mathcal S_{tg}$ and
$(\pbpg)=\mathcal S_{bg}$. Also, we use the following 
expressions :
\bea\label{rhopol-un}
\mathcal F_{1}&=&\nonumber\frac{g^2 g_s^2}{24\stg^2\sbg^2}\Bigg[\left\{\lvert \fla\rvert^2\left(1+\frac{m_t^2}{2m_W^2}\right)
+3\mathrm {Re}\fla \mathrm {f_{2R}^*}\frac{m_t}{m_W}
+3|\frb|^2\frac{m_t^2}{2m_W^2}\right\}
\\\nonumber
&\times&\Big\{\left[\stb+\stg-\sbg\right]\left[2\stg\stb\sbg-m_t^2\sbg^2\right]-\stg\sbg\left[\sbg^2-\stg^2\right]\Big\}\\\nonumber
&-&\lvert \fla\rvert^2\frac{m_t^2}{m_W^2}\stg^2\sbg^2
+\mathrm {Re}\fla \mathrm {f_{2R}^*}\frac{m_t}{m_W}\Big\{2\stg\sbg^2[\sbg-\stg]\Big\}\\\nonumber
&+&\frac{\lvert\frb\rvert^2}{m_W^2}\Big\{m_t^2[-\sbg^2(\sbg\stg-\stb^2+\sbg^2)]+\stg\sbg\big[\stb(\stg+\stb-\sbg)\\\nonumber
&\times&\{3(\sbg-\stg)-\stb\}+\{(\sbg+\stg)(\sbg^2-\stg^2)-\stg^3\}\big]\Big\}\Bigg]
\eea
\bea\label{rho-pol}
\mathcal F_{2}&=&\nonumber\frac{g^2 g_s^2}{12\stg^2\sbg^2}\Big[\left\{\lvert\fla\rvert^2\left(1-\frac{m_t^2}{2m_W^2}\right)
+\mathrm {Re}\fla \mathrm {f_{2R}^*}\frac{m_t}{m_W}+|\frb|^2\frac{m_t^2}{2m_W^2}\right\}m_t\sbg\\\nonumber
&\times&\left[\left(\sbn+\sgn\right)\stg\left(\stb+\stg\right)-\sbg\left(\stb\sgn+\stg\sbn+m_t^2\sbn-\sgn\right)
+\stb\stg\sbn\right]\\\nonumber
&-&|\fla|^2\frac{m_t^2}{m_W^2}\stg\sbg^2\sgn
+\mathrm {Re}\fla \mathrm {f_{2R}^*}\frac{1}{m_W}\Big\{2\stg^2\sbg[\stb\sgn-\stg\sbn]\Big\}\\\nonumber
&+&|\frb|^2\Big\{\sbn(\stb-\sbg+\stg)(m_t^2\sbg^2-3\sbg\stg^2+\sbg^2\stg-2\sbg\stg\stb)\\\nonumber
&+&2\sbg\stg^2\stb+\sgn(\stb-\sbg-\stg)(-\sbg\stg\stb+\sbg\stg^2-\sbg^3)-2\sbg\stg^3\\\nonumber
&+&\sbg^2(\stb\sbg+2\sbg\stg+\sbg^2+m_t^2\stg)\Big\}\Big]
\eea

The diagonal elements of the spin density matrix for single-top production in $tW$ channel can be written as 
\beq
\rho(\pm,\pm)=\mathcal{F}_1\pm\mathcal{F}_2
\eeq
with $\sgn=(p_g.n_3)$ and $\sbn=(p_b.n_3)$ and the off-diagonal elements are 
\beq
\rho(\pm,\mp)=\mathcal{F}_2 
\eeq
with $\sgn=(p_g.(n_1\pm i n_2))$ and $\sbn=(p_b.(n_1\pm i n_2))$ where $n^{\mu}_{i}$'s ($i=1,2,3$) are the 
four-vectors with the properties 
$n_i \cdot n_j =-\delta_{ij}$ and $n_i \cdot p_t =0$. $n_1$, $n_2$ and
$n_3$ represent spin four-vectors of the top quark with spin
respectively along the $x$, $y$ and $z$ axis in the top rest frame.

 \bibliographystyle{apsrev4-1.bst}

\end{document}